\documentclass[useAMS,usenatbib]{mn2e}
\usepackage{journal_abbreviations}
\usepackage{graphicx, amsmath, amssymb}
\usepackage{enumerate}
\usepackage{float}

\usepackage{mathrsfs}
\usepackage{graphics}

\title[Variable initial mass function from ATLAS$^{\mathit{3D}}$]{An assessment of the evidence from ATLAS$^{\rm{3D}}$ for a variable initial mass function}
\author[Clauwens et al.]{Bart Clauwens$^{1,2}$\thanks{E-mail: clauwens@strw.leidenuniv.nl},
Joop Schaye$^{1}$, Marijn Franx$^{1}$\\
$^{1}$Leiden Observatory, Leiden University, PO Box 9513, 2300 RA Leiden, The Netherlands\\$^{2}$Instituut-Lorentz for Theoretical Physics, Leiden University, 2333 CA Leiden, The Netherlands\\}

\begin{document}

\date{Accepted, 17 March 2015. Received 17 March 2015; in original form, 3 June 2014}
%\date{2015 Mar 16}

\pagerange{\pageref{firstpage}--\pageref{lastpage}}% \pubyear{2013}

\maketitle

\label{firstpage}

\begin{abstract}

The ATLAS$^{\rm{3D}}$ Survey has reported evidence for a non-universal stellar initial mass function (IMF) for early type galaxies (ETGs) \citep{Cappellari12, ATLAS15, ATLAS20}. The IMF was constrained by comparing stellar mass measurements from kinematic data with those from spectral energy distribution (SED) fitting. Here we investigate possible effects of scatter in the reported stellar mass measurements and their potential impact on the IMF determination. We find that a trend of the IMF mismatch parameter with the kinematic mass to light ratio, comparable to the trend observed by \citet{Cappellari12}, could arise if the Gaussian errors of the kinematic mass determination are typically 30\%. Without additional data, it is hard to separate between the option that the IMF has a true large intrinsic variation or the option that the errors in the determination are larger than anticipated. A correlation of the IMF with other properties would help to make this distinction, but no strong correlation has been found yet. The strongest correlation is with velocity dispersion. However, it has a large scatter and the correlation depends on sample selection and distance measurements. The correlation with velocity dispersion could be partly caused by the colour-dependent calibration of the surface brightness fluctuation distances of \citet{Tonry01}. We find that the K-band luminosity limited ATLAS$^{\rm{3D}}$ Survey is incomplete for the highest M/L galaxies below $10^{10.3} {\rm M_{\odot} }$. There is a significant IMF - velocity dispersion trend for galaxies with SED masses above this limit, but no trend for galaxies with kinematic masses above this limit. We also find an IMF trend with distance, but no correlation between nearest neighbour ETGs, which excludes a large environmental dependence. Our findings do not rule out the reported IMF variations, but they suggest that further study is needed.
\end{abstract}


\section{Introduction}
\label{SectionIntroduction}

The stellar initial mass function (IMF)  has historically been assumed to be universal, in the sense that it does not depend on environment. The IMF was assumed to be independent of galaxy age, galaxy type, metallicity or any other astrophysical variable, with the possible exception of population III stars and stars forming near the galactic center e.g. \citep{Kroupa13}. Since the exact mechanisms that cause the formation of stars of varying masses from an initial cloud of gas and dust are not well understood, the assumption of the universality of the IMF is partially motivated by a desire for simplicity, but it is also supported by direct measurements of stellar mass distributions in our immediate vicinity e.g. \citep{Chabrier03,Kroupa13,Bastian11, Kirk11}. It is reasonable to assume that the IMF does differ in more extreme environments, but this is hard to measure directly.

On a galactic scale, evidence has recently been found in favour of a non-universal IMF for early type galaxies (ETGs), typically depending on the velocity dispersion of the galaxy. The evidence comes partly from differing spectral features of low- and high-mass stars \citep{Barbera13,Dokkum12,Conroy12,Pastorello14} and partly from mass measurements of stellar systems via strong gravitational lensing \citep{Treu10,Brewer12,Oguri14,Barnabe13} or the modeling of stellar kinematics \citep{Conroy13, Tortora13, Dutton13, Cappellari12,ATLAS15,ATLAS20}. However, the nearest known strong lens provides conflicting evidence \citep{Smith13} and a recent study of the low mass X-ray binary population in eight ETGs also points towards a universal IMF \citep{Peacock14}. \citet{Conroy13} find good agreement between IMF variations from spectral features and from kinematics for stacks of galaxies. On the other hand, a recent comparison between dynamical and spectroscopic results by \citet{Smith14} shows that the IMF measurements of \citet{Conroy12} and those of \citet{ATLAS20} agree only superficially and not on a galaxy by galaxy basis. Also, a recent detailed spectral analysis of three nearby ETGs by \citet{MartinNavarro14} found at least one massive galaxy (NGC4552) for which the IMF varies strongly with radius from the centre.

Estimating the IMF via a mass measurement independent of the spectral features has the obvious disadvantage that it is only sensitive to the overall missing mass, which could be a superposition of low-mass stars, stellar remnants and dark matter. The advantage is, however, that the measurement is independent of broad-band SED fitting or the fitting of specific gravity sensitive spectral lines and therefore it can either confirm or refute IMF trends that might be deduced from the intricacies of integrated spectra of galaxies. Gravitational lensing has the disadvantage that it is a mass measurement along a cylinder and therefore is relatively sensitive to dark matter or any other matter along the line of sight. A potentially cleaner way to obtain a mass estimate of only the baryonic matter, is to analyze the kinematics of the central parts of ETGs, whose mass is believed to be dominated by baryons.

An attempt to observe and explain the stellar kinematics in the central regions of ETGs has been undertaken by the ATLAS$^{\rm{3D}}$ Survey \citep{ATLAS1}. The aim of this survey has been to obtain integral field spectroscopy with SAURON \citep{Bacon01} of all 260 ETGs with mass approximately greater than $6 \times10^{9} {\rm {\rm M_{\odot} }}$  that are within 42 Mpc distance from us in the northern hemisphere. This volume-limited sample yields a large collection of kinematic data, which has been used, among other things, to estimate the stellar masses of these galaxies. Comparing these kinematic measurements with the stellar masses measured by fitting the SEDs with stellar population synthesis models, provides a direct probe of the IMF normalization in these galaxies. A clear trend of IMF normalization with velocity dispersion or with mass-to-light ratio has been reported by \citet{Cappellari12,ATLAS15,ATLAS20}, resulting in: (I) A Chabrier-like normalization at low mass-to-light ratios, which agrees with the one inferred for spiral galaxies, (II) A Salpeter normalization at larger $(M/L)$ consistent, on average, with some results from strong lensing and (III) a normalization more massive than Salpeter for some of the galaxies with high $(M/L)$ broadly consistent with measurements of spectral features in massive galaxies that indicate a substantial population of dwarf stars \citep{Cappellari12}.

This article consists of a critical review of some of the methods and results from the ATLAS$^{\rm{3D}}$ Survey. Section \ref{SectionAtlas} introduces the ATLAS$^{\rm{3D}}$ Survey and the JAM method used to fit the kinematical data. In section \ref{SectionEvidence} the evidence from \citet{Cappellari12} for a non-universal IMF is investigated. Specifically it is shown that the large reported trend between the kinematic mass to light ratio and the IMF mismatch parameter, interpreted as an effect of IMF variations, could also be caused by measurement errors in the kinematic mass of the order 30\%. Section \ref{SectionCorrelations} presents correlations of the IMF normalization with astrophysical variables. Section \ref{sectionGSMF} shows that the effect of the non-universal IMF implied by the original ATLAS$^{\rm{3D}}$ analysis on observations of the Galaxy Stellar Mass Function (GSMF) at higher redshift is small. Also the stellar mass completeness limit of the ATLAS$^{\rm{3D}}$ Survey is shown to be $10^{10.3}{\rm {\rm M_{\odot} }}$. In section \ref{SectionCompleteness} we demonstrate that the inferred systematic IMF trend with velocity dispersion is dependent on the precise selection cut that is made at the low mass end. In particular we show that this trend is virtually absent for the mass complete sample of galaxies with kinematic stellar masses larger than $10^{10.3}{\rm {\rm M_{\odot} }}$. In section \ref{SectionDistance} we show that the systematic variation of the IMF with velocity dispersion is accompanied by a systematic variation with distance. This could be interpreted as a genuine effect of the cosmic environment on the IMF, but more probably it points towards biases in the used distance catalog which, as a side-effect, show up as a dependence of the IMF on the velocity dispersion of an ETG. Part of the IMF trend can be attributed to colour-dependent calibration issues of the surface brightness fluctuation (SBF) distance measurements and we show that the IMF trend is absent for galaxies at a distance larger than 25 Mpc.  Finally, we summarize our conclusions in section \ref{SectionConclusion}.

\section{The ATLAS$^{\rm{3D}}$ Survey}
\label{SectionAtlas}

The ATLAS$^{\rm{3D}}$ project improves on previous studies in two ways. On the one hand the number of observed objects, 260, is much larger than before. On the other hand, progress has been made in modeling the observed stellar dynamics. The  ATLAS$^{\rm{3D}}$ team's Jeans Anisotropic Multi-Gaussian Expansion (JAM) modeling method is introduced in \citet{Cappellari08,Cappellari12*}. The JAM method uses the minimum number of free parameters that are needed to fit the integral field observations. It assumes axisymmetry for all galaxies, with the inclination $i$ as a free parameter. The mass-to-light ratio $\Upsilon$ is assumed to be the same throughout the whole observed region, but it can vary from galaxy to galaxy. The conversion of the observed luminosity density to a matter density depends on $i$ and $\Upsilon$ and is done with the multi Gaussian Expansion (MGE) parameterization of \citet{Emsellem94}.

The JAM method consists of solving the Jeans equations, with the extension (with respect to the isotropic case) of an orbital anisotropy parameter $\beta_z$. The velocity ellipsoid is assumed to be aligned at every position in the galaxy with the cylindrical coordinates $(R,z)$ and the ratio between the two axes of this ellipsoid is assumed to be the same within the central part of the galaxy, leaving one extra free parameter, $\beta_z=1- \overline{{v_z}^2}/\overline{{v_R}^2}$. Although the velocity ellipsoid will in reality be more complicated, this simple $\beta_z$ parameter suffices to connect the model to the observations. Apart from the three parameters $i$, $\Upsilon$ and $\beta_z$, six different parameterizations of the dark matter halo are used, but the main conclusions are found to be insensitive to dark matter, because for all six halo parameterizations the kinematics of the central part of the ETGs are dominated by baryonic matter.

As shown in \citet{Cappellari12}, this model not only suffices to fit the integral field spectroscopic observations, it also puts very tight constraints on the $\Upsilon$ parameter. It is this feature that makes it possible to measure the IMF normalization, but let us first take a quick look at the other two free parameters.

The main argument in favour of the model is the fact that it is able to reproduce the integral field spectroscopy of a complete and very diverse set of galaxies using only a small number of free parameters. However this same argument also works against it, because \citet{Cappellari08} note that for galaxies observed at low inclination, the lowest $\chi$-squared fit is often obtained for an unrealistic set of parameter values, because of a degeneracy between $i$ and $\beta_z$. The model prefers too high values for $i$ and too low values for $\beta_z$. Restricting the anisotropy to a flat ellipsoid, $\beta_z>0.05$ as observed for edge-on galaxies, does remove the degeneracy, but this example shows that a good fit does not necessarily prove that the model corresponds to physical reality.

\begin{figure}
\center
\includegraphics[width=1.0\columnwidth]{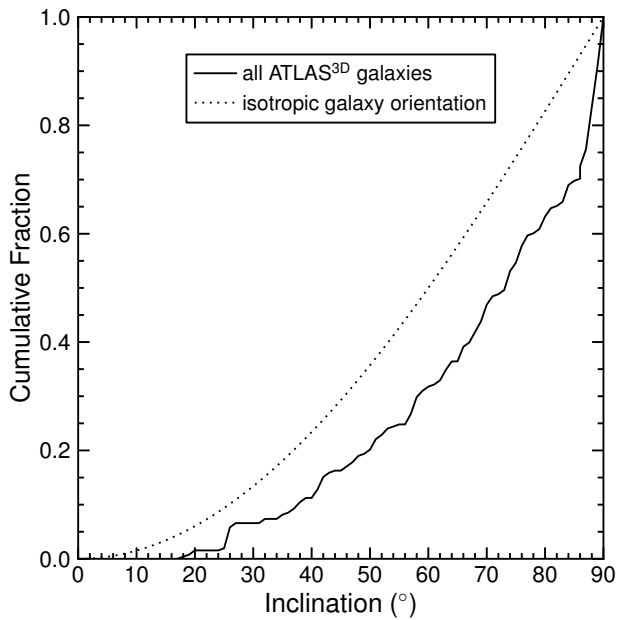}
\caption{Distribution of the JAM model inclinations of all the ATLAS$^{\rm{3D}}$ ETGs compared to an isotropic distribution of inclinations. The Kolmogorov-Smirnov statistic for this comparison is 0.22 with a corresponding probability $p < 10^{-11}$.}
\label{GraphInclination}
\end{figure}

Because of the large size of the survey we can look at the distribution of inclinations. Figure \ref{GraphInclination} compares the observed distribution of inclinations with that expected for randomly oriented galaxies. The Kolmogorov-Smirnov statistic for this comparison is 0.22 with a corresponding probability $p < 10^{-11}$. With respect to the isotropic case there is a shortage of $\sim20\%$ of galaxies with inclinations smaller than $45^\circ$ and an excess of $\sim20\%$ of galaxies with inclinations larger than $85^\circ$. This either indicates that the model still has a tendency to overestimate the inclination or that the ETGs in our local neighborhood are preferably aligned with our line of sight. In principle a measurement error in the inclination could result in an error in the determined IMF mismatch parameter. A priori there is no clear reason to assume that this would not bias the determination of the IMF. However there is no significant correlation (Pearson $R^2=0.01$) between inclination and the IMF mismatch parameter, lending an a posteriori credibility to the retrieved IMF normalization\footnote{However the fact that the five galaxies with the lowest IMF mismatch parameter all have an inclination larger than $85^{\circ}$ suggests that at least for these galaxies the true inclination might be smaller, or the IMF mismatch parameter dependent on the assumed inclination.}. In the following section we will take a detailed look at the predictions for the mass-to-light ratio $\Upsilon$ and the implications for the IMF normalization.

\section{The ATLAS$^{\rm{3D}}$ evidence for a non universal IMF}
\label{SectionEvidence}

The precision with which deviations from universality in the IMF can be measured, depends on the errors in the two independent measurements of $(M/L)$\footnote{The $(M/L)$ and luminosity measurements in this paper refer to the r-band, as is the case for the ATLAS$^{\rm 3D}$ papers.} from respectively SED fitting and the stellar kinematics via the JAM method. $(M/L)_{\rm SED}$\footnote{The ATLAS$^{\rm{3D}}$ papers denote this variable as $(M/L)_{\rm Salp}$. We will refer to it as $(M/L)_{\rm SED}$ in this paper.} is obtained by using the spectral fitting models of \citet{Vazdekis12}, with standard lower and upper mass cut-offs for the Salpeter IMF of $0.1$ M$_{\odot}$ and $100$ M$_{\odot}$. A comparison has been made with the $(M/L)$ values from \citet{Conroy12}, who use an independent set of spectra spanning a longer wavelength range and a different stellar population synthesis model. For the set of 35 galaxies that are present in both studies, the differences between the two $(M/L)$ measurements are consistent with an error per galaxy per measurement of 6\%, which suggests that $(M/L)_{\rm SED}$ is quite robust \citep{ATLAS20}.

By comparing predictions from models with different dark matter halos, \citet{ATLAS15} estimate the JAM modeling errors in $(M/L)_{\rm kin}$ to be 6\%. We will use $(M/L)_{\rm kin}$\footnote{The ATLAS$^{\rm{3D}}$ papers denote this variable as $(M/L)_{\rm stars}$. We will refer to it as $(M/L)_{\rm kin}$ in this paper.} to denote the stellar mass-to-light ratio of the best fit JAM model with a NFW dark matter halo with a fitted virial mass $M_{200}$, also referred to as model B by \citet{Cappellari12}, where $M_{200}$ denotes the mass of a 200 times overdensity dark matter halo. Galaxies with a clear bar structure give lower quality fits than galaxies with no bars. Apart from this there may be errors from distance measurements and from photometry.

\begin{figure*}
\includegraphics[width=1.0\columnwidth]{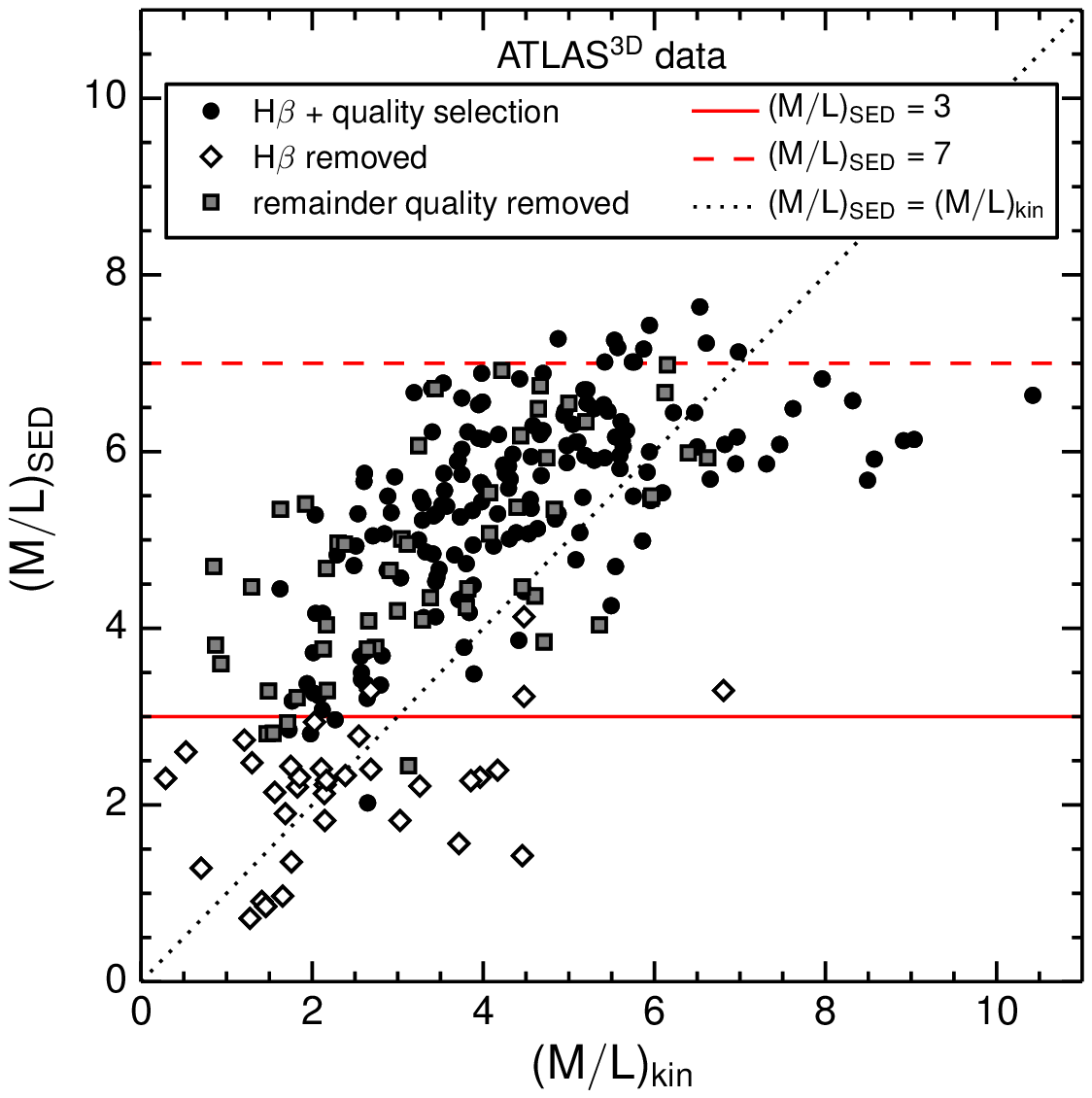}
\includegraphics[width=1.0\columnwidth]{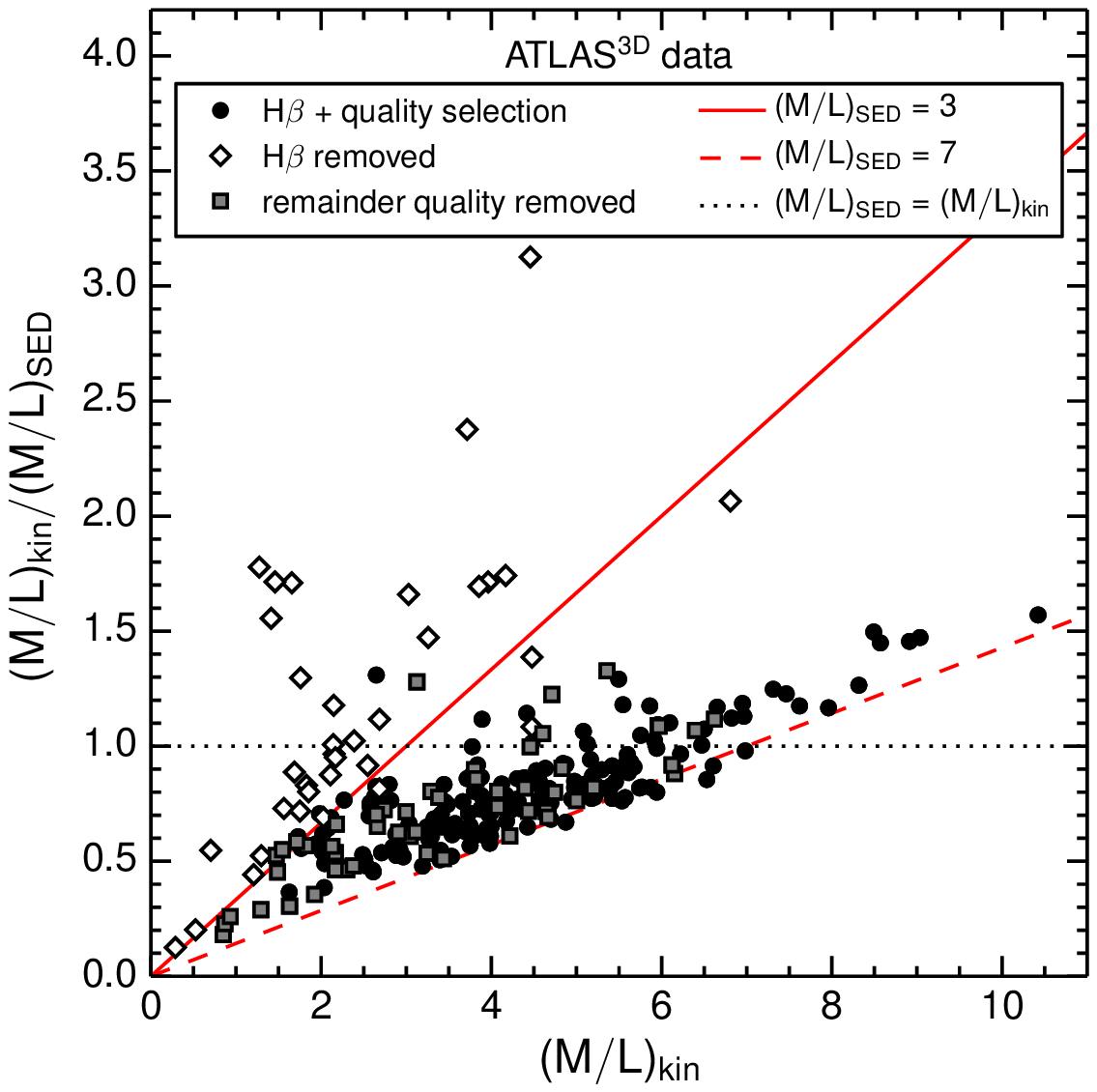}
\includegraphics[width=1.0\columnwidth]{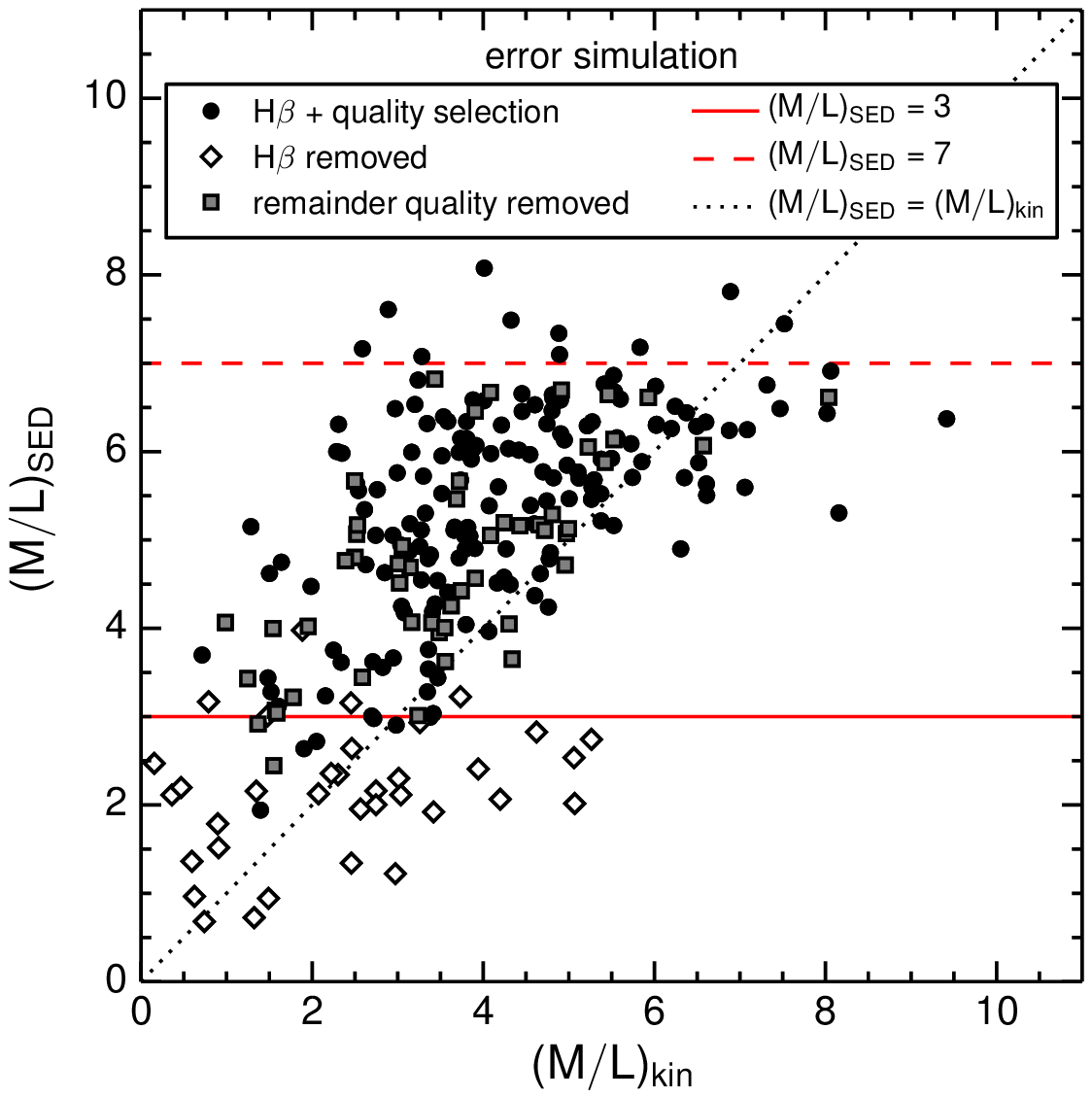}
\includegraphics[width=1.0\columnwidth]{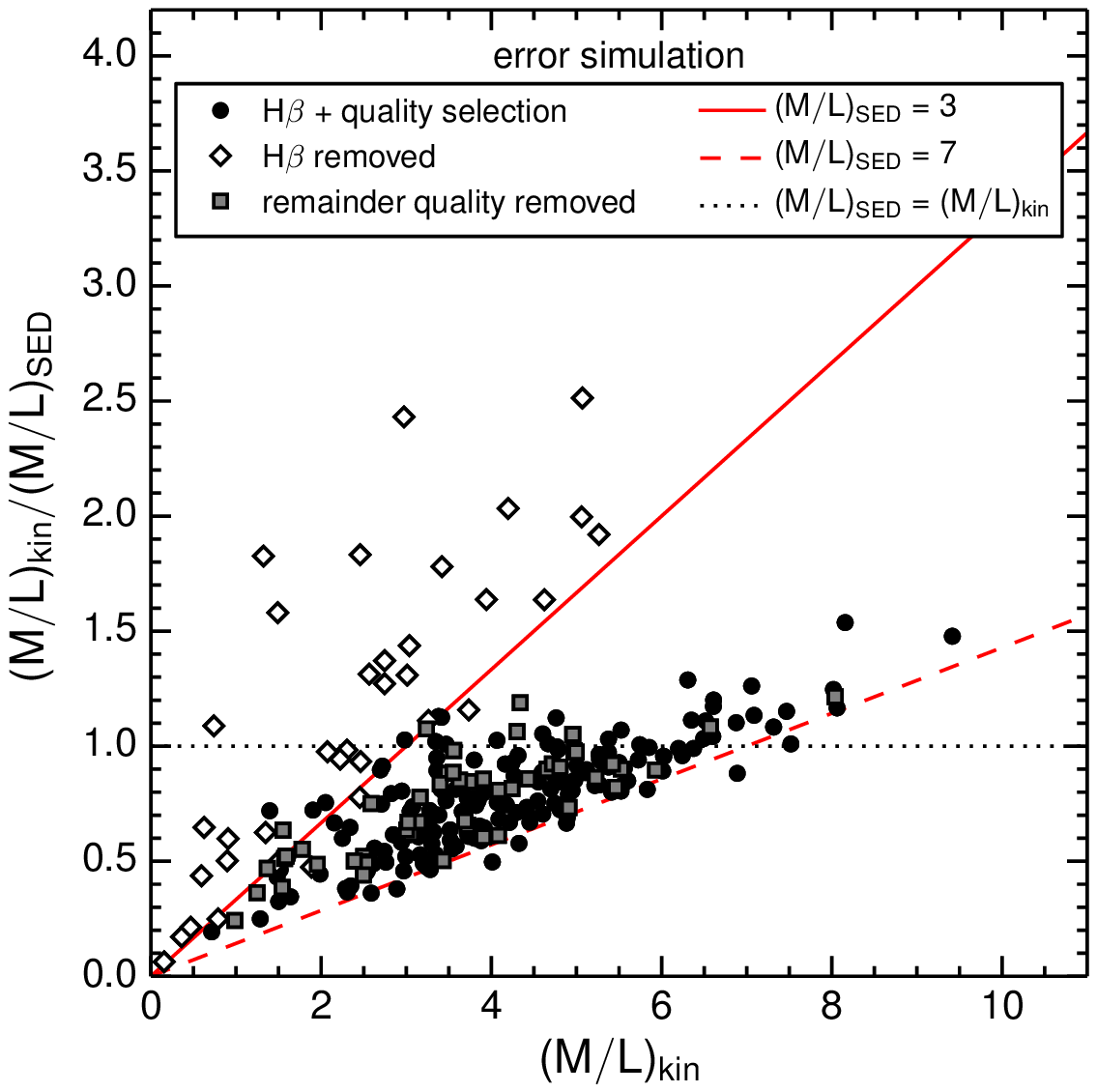}
\caption{Top left panel: comparison of the JAM model stellar mass-to-light ratio, $(M/L)_{\rm kin}$, with the ratio inferred from Stellar Population Synthesis SED fits assuming a Salpeter IMF, $(M/L)_{\rm SED}$, for the ATLAS$^{\rm{3D}}$ dataset. Open diamonds indicate galaxies with a young stellar population, selected by having H$\beta$ absorption with an equivalent width larger than $2.3$ \AA. These galaxies tend to have strong radial gradients in their population which makes both $(M/L)$ determinations uncertain \protect\citep{Cappellari12}. This selection is almost identical to selecting all galaxies with $(M/L)_{\rm SED} < 3$ (horizontal solid line). Grey squares indicate the remaining galaxies with a (quality = 0) label, meaning: ``either inferior data quality (low S/N) or a problematic model (e.g. due to the presence of a strong bar or dust, or genuine kinematic twists).'' Black circles are the remaining high-quality galaxies. The horizontal dashed line at $(M/L)_{\rm SED} \approx 7$ denotes the theoretical maximum for a simple stellar population of the age of the universe with a Salpeter IMF; Top right panel: the ``IMF mismatch parameter'', i.e. the ratio $(M/L)_{\rm kin}/(M/L)_{\rm SED}$, as a function of $(M/L)_{\rm kin}$. This plot is similar to the upper middle panel of Figure 2 from \protect\citet{Cappellari12} apart from the selection of galaxies and a logarithmic axis; The bottom panels show the same plots for simulated data for which it is assumed that there are no intrinsic IMF variations (within the black and grey data points), but for which the perceived variations are caused by a random Gaussian errors of 6\% in $(M/L)_{\rm SED}$ and 29.9\% in $(M/L)_{\rm kin}$. The black and grey data points are also renormalised by a factor of 0.785, see Table \ref{TableMismatch}. The error of 29.9\% is chosen such that the standard deviation in the mismatch parameter in the error simulation is exactly the same as in the ATLAS$^{\rm{3D}}$ data. Both the qualitative as the quantitative behaviour are reproduced pretty well. The Pearson $R^2$ for the black and grey points of the right panels is $0.674$ for the data and $0.605 \pm0.040$ for 10.000 runs of the Gaussian error simulation (for the specific run that is shown here it is 0.630). The white diamonds require a larger normalisation of 1.192 and error of 51.2\%.}
\label{GraphStarSalp}
\end{figure*}

\begin{figure*}
\includegraphics[width=1.0\columnwidth]{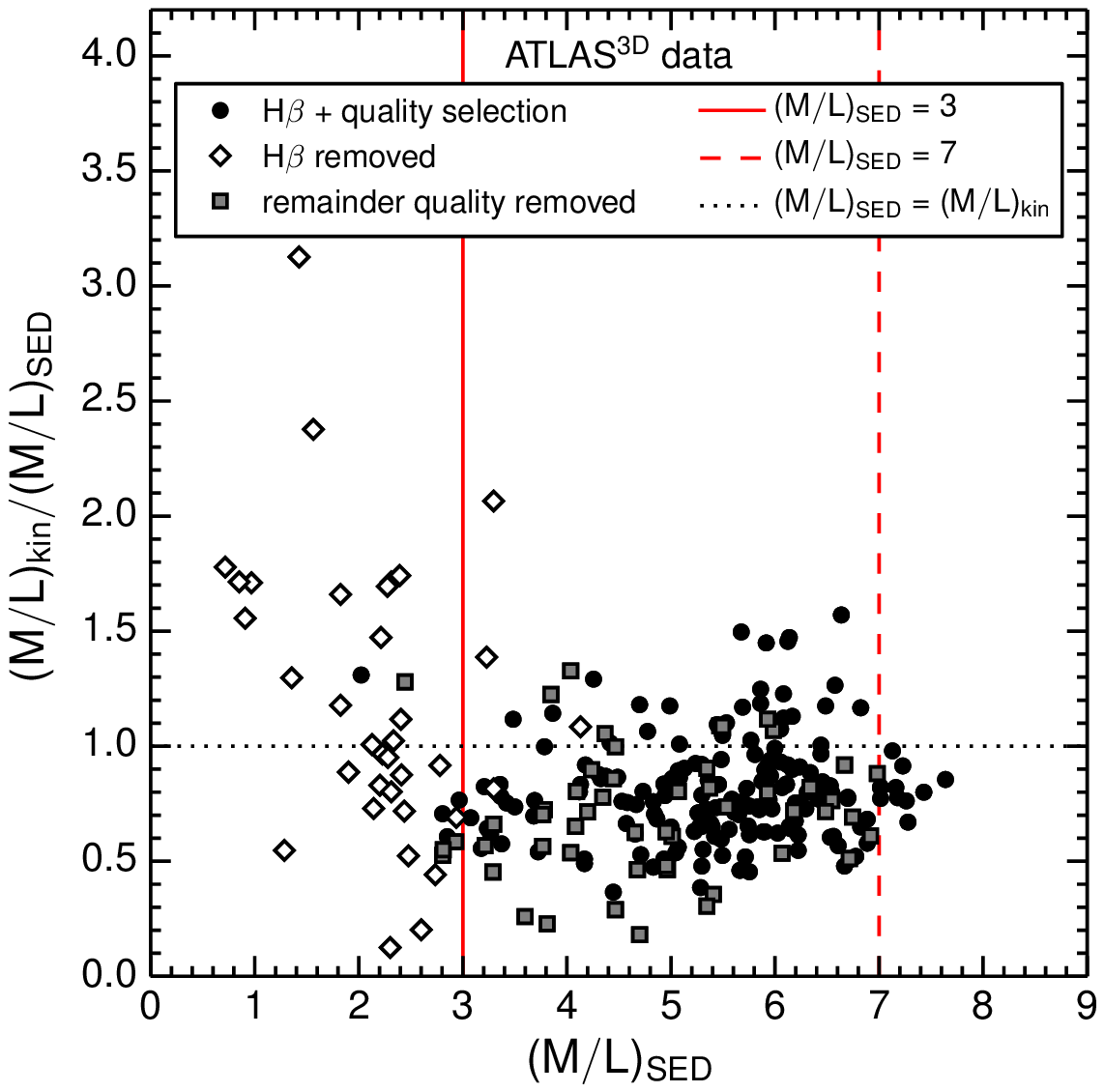}
\includegraphics[width=1.0\columnwidth]{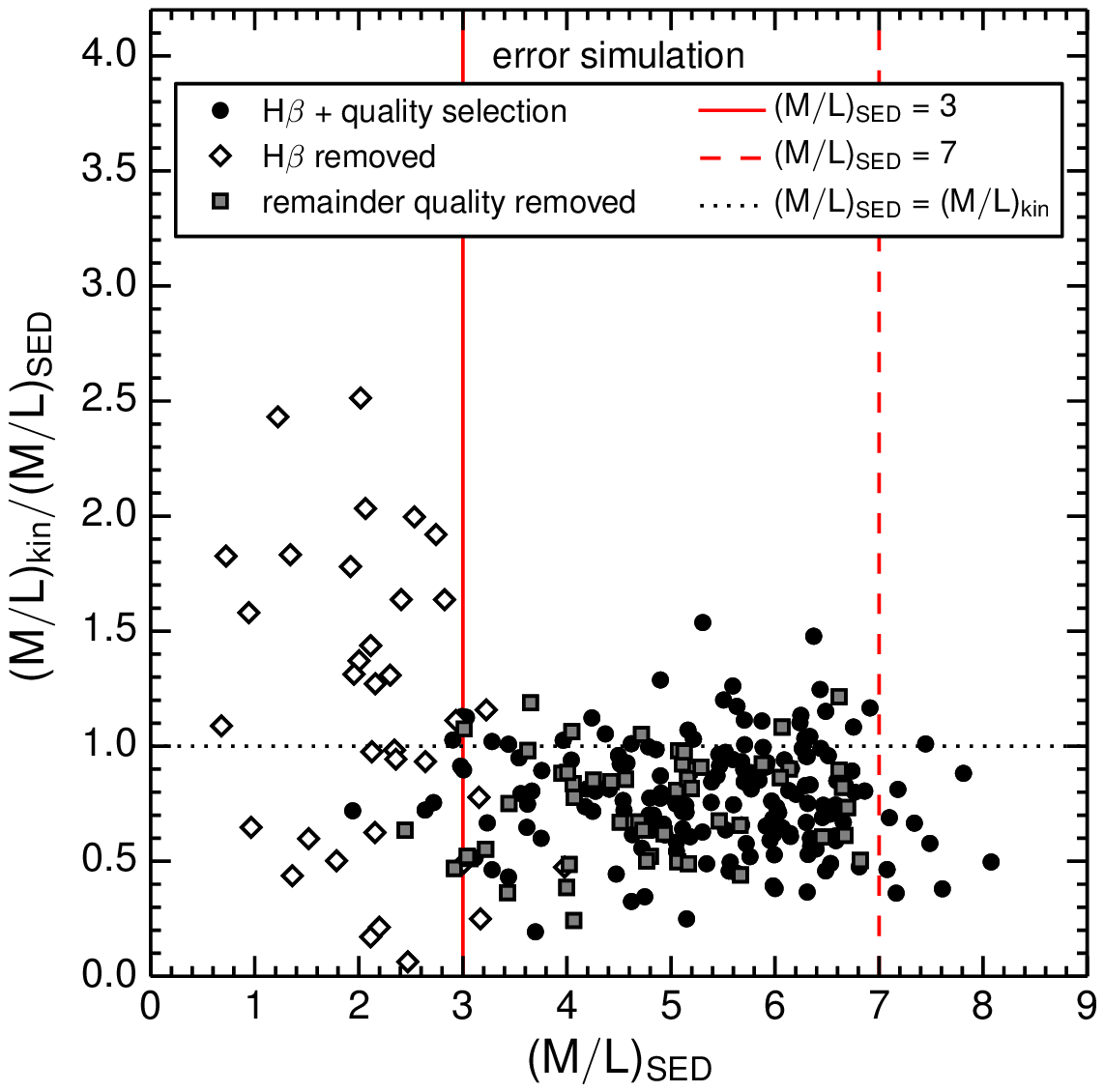}
\caption{For the same data as Figure \ref{GraphStarSalp} this shows the dependency of the IMF mismatch parameter on  $(M/L)_{\rm SED}$. Both the ATLAS$^{\rm{3D}}$ data and the error simulation show a negligible correlation for the black and grey data points, with a Pearson $R^2$ of 0.02 for the data and 0.00 for the simulation.}
\label{GraphExtra}
\end{figure*}

Figure \ref{GraphStarSalp} (top panels) compares the two types of $(M/L)$ determinations from the ATLAS$^{3D}$   Survey. Clearly, $(M/L)_{\rm SED}$ and $(M/L)_{\rm kin}$ do not agree within the 6\% error associated with the $(M/L)_{\rm SED}$ determination and the 6\% JAM model error. The difference could be due to a systematic IMF trend, random variations in the IMF, distance measurement errors and photometry errors. Our aim is to better understand these effects.

\citet{Cappellari12} present the ATLAS$^{\rm{3D}}$ results in a way analogous to Figure \ref{GraphStarSalp} (top right panel), without the open diamond symbols. One should be cautious drawing conclusions about the IMF from the correlation in this graph between $(M/L)_{\rm kin}$ and the ``IMF mismatch parameter''  $\alpha \equiv [(M/L)_{\rm kin}] / [(M/L)_{\rm SED}]$ for three reasons. Firstly, galaxies with still ongoing star formation (selected by having H$\beta$ absorption with an equivalent width larger than 2.3 \AA) generally have a strong radial gradient in their stellar population. This makes both $(M/L)$ determinations uncertain, which is the reason why they are excluded from the analysis by \citet{Cappellari12}. This does, however, induce an unavoidable bias. Figure \ref{GraphStarSalp} (top left panel) shows that this H$\beta$ selection is almost equivalent to removing all galaxies with $(M/L)_{\rm SED}<3$. Figure \ref{GraphStarSalp} (top right panel) shows that this creates an ``upper zone of avoidance'' which strengthens the correlation between $\alpha$ and $(M/L)_{\rm kin}$.

Secondly, $(M/L)_{\rm SED}$ is not a pure measurement. It is a fit of measurements to a Salpeter stellar synthesis model and hence it does not have Gaussian random error behaviour. More specifically, there is a clear theoretical maximum value of $(M/L)_{\rm SED} \approx 7$ which corresponds to a simple stellar population  of the age of the universe with a Salpeter IMF. Regardless of any errors in SED-fitting, JAM-modeling, distance measurements and photometry, this maximum will always be respected. As can be seen in Figure \ref{GraphStarSalp} (top right panel), this constitutes a ``lower zone of avoidance'' which is actually responsible for most of the correlation.

Thirdly, and not completely independent of the previous two points:  any error in the kinematic $(M/L)$ determination will show up as a radial scatter which emanates from the origin in Figure \ref{GraphStarSalp} (top right panel) and may thus induce a spurious correlation.

\begin{table*}
\begin{center}
\begin{tabular}{llllll}
\hline
galaxy selection & $\overline{\alpha}$ & $\sigma(\alpha)$ & $\sigma(\alpha)/\overline{\alpha}$ & number of galaxies \\   \hline
H$\beta$ + quality selection (black circles) & 0.808 & 0.226 & 28.0\% & 171 \\ 
H$\beta$ removed (white diamonds) & 1.192 & 0.615 & 51.6\% & 35 \\ 
remainder quality removed (grey squares) &  0.710 & 0.265 & 37.3\% & 52 \\ 
(black circles + grey squares) & 0.785 & 0.239 & 30.5\% & 223 \\   \hline
\end{tabular}
\caption{Average IMF mismatch parameter $\overline{\alpha}$, the standard deviation $\sigma(\alpha)$, the relative standard deviation $\sigma(\alpha)/\overline{\alpha}$ and the number of galaxies in the selection for the galaxy samples corresponding to different selection methods as used in Figure \ref{GraphStarSalp} and other Figures throughout this article.}
\label{TableMismatch}
\end{center}
\end{table*}

In order to assess to what extent the upper right panel of Figure \ref{GraphStarSalp} alone, or equivalently the upper middle panel of Figure 2 from \citet{Cappellari12}, constitutes convincing evidence for IMF variations, we simulate the effect of Gaussian random errors in both the determination of $(M/L)_{\rm{SED}}$ and $(M/L)_{\rm{kin}}$ on this figure. Assuming no intrinsic IMF variations, these errors will lead to an expected scatter in the perceived IMF mismatch parameter. We fix the Gaussian errors in $(M/L)_{\rm{SED}}$ to the reported value of 6\%, but use a Gaussian error of 29.9\% in $(M/L)_{\rm{kin}}$, which represents the total error in the kinematic mass-to-light determination, including a JAM modelling error (reported at 6\%), errors from photometry and errors from the distance determination, which will be discussed at length in section \ref{SectionDistance}. The value of 29.9\% is chosen such that the kinematic and SED errors together combine to give the 30.5\% scatter found in the data for all galaxies that have not been rejected because of H$\beta$ absorption, see Table \ref{TableMismatch}. The question now is whether these random errors can produce at the same time a relation between $(M/L)_{\rm{kin}}$ and $\alpha$ similar to that in Figure \ref{GraphStarSalp} upper right panel.

Figure \ref{GraphStarSalp} (lower panels) shows the results of the error simulation for data with no intrinsic IMF variations. For all the galaxies that have not been rejected on basis of H$\beta$ absorption we simulate a random value for $(M/L)_{\rm{kin}}$ based on the observed value of $(M/L)_{\rm{SED}}$ from ATLAS$^{\rm{3D}}$ multipled by the average normalisation of 0.785 (see Table \ref{TableMismatch}) and we add a random Gaussian error of 29.9\%. Hereafter we add a 6\% random Gaussian error to $(M/L)_{\rm{SED}}$. For the H$\beta$ removed galaxies we use a normalisation of 1.192 and respective errors of 51.2\% and 6\%. As can be seen in the lower panels of Figure \ref{GraphStarSalp} the data from simulated errors looks very similar to that from the real ATLAS$^{\rm{3D}}$ measurements. Especially we retrieve the strong trend of the IMF mismatch parameter with $(M/L)_{\rm{kin}}$. However the correlation of this trend in the real data (Pearson $R^2= 0.674$ for the combined black circles and grey squares) is higher than that in most of the error simulations (Pearson $R^2=0.605 \pm 0.040$). This 1.7$\sigma$ deviation could indicate that Gaussian errors alone are not enough to explain the observed trend between $(M/L)_{\rm{kin}}$ and $\alpha$, although the significance of this is limited and non-Gaussianities in the errors are likely to increase this correlation. Figure \ref{GraphExtra} shows that also the relation between $(M/L)_{\rm{SED}}$ and $\alpha$ is well reproduced by the error simulation. The data has a Pearson $R^2$ of 0.02 versus 0.00 in the simulation. A negative correlation could have been the result of hypothetical large measurement errors in $(M/L)_{\rm{SED}}$.

These issues do not definitely imply that the observed trend is caused by errors. For the sake of the argument, true Gaussian IMF variations would look exactly the same as Gaussian measurement errors in $(M/L)_{\rm{kin}}$. It does show however that it is hard to draw conclusions based solely on the correlation between $\alpha$ and $(M/L)_{\rm kin}$. It is important to look for accompanying correlations of the IMF mismatch parameter $\alpha$ with different variables, not only to find the physical processes that might explain the trend, but also to rule out that the trend is a result of the complicated interplay between the selection effects and the different measurement and model errors.

Even in the extreme case when the variations of the IMF mismatch parameter $\alpha$ within the the ATLAS$^{\rm{3D}}$ Survey would be completely due to errors, the average value of $\alpha$ from Table \ref{TableMismatch} can still be compared with determinations of the IMF by different studies, as alluded to in the introduction. This average normalization for the ATLAS$^{\rm{3D}}$ ETGs is different from the Chabrier IMF which holds for our galaxy. However when comparing to other studies one has to take into account the unknown systematics of comparing different IMF determination methods. This is beyond the scope of this work. We will focus solely on the evidence for IMF variations present within the ATLAS$^{\rm{3D}}$ Survey.

\section{Correlations with the IMF mismatch parameter}
\label{SectionCorrelations}

In the previous section we confirmed that at face value the ATLAS$^{\rm{3D}}$ data suggests a non-universal IMF. The robustness of this outcome however critically depends on the size of the assumed modelling and measurement errors in the kinematic mass determination. For this reason, it would be good to find some independent correlation of the IMF mismatch parameter with some other observable in order to convince ourselves of the robustness of this result. Moreover, correlations are to be expected within any theoretical model for IMF variations. The IMF could for example correlate with the age of the galaxy through a dependence on redshift, it could be related to the mass of the galaxy via gas recycling, the pressure of the interstellar matter or the intensity of star formation, it could depend on the galaxy metallicity or it could be influenced by the cosmic environment etc. Any correlation could also point the way to an understanding of the underlying physical mechanisms.

\begin{figure*}
\includegraphics[width=1.0\columnwidth]{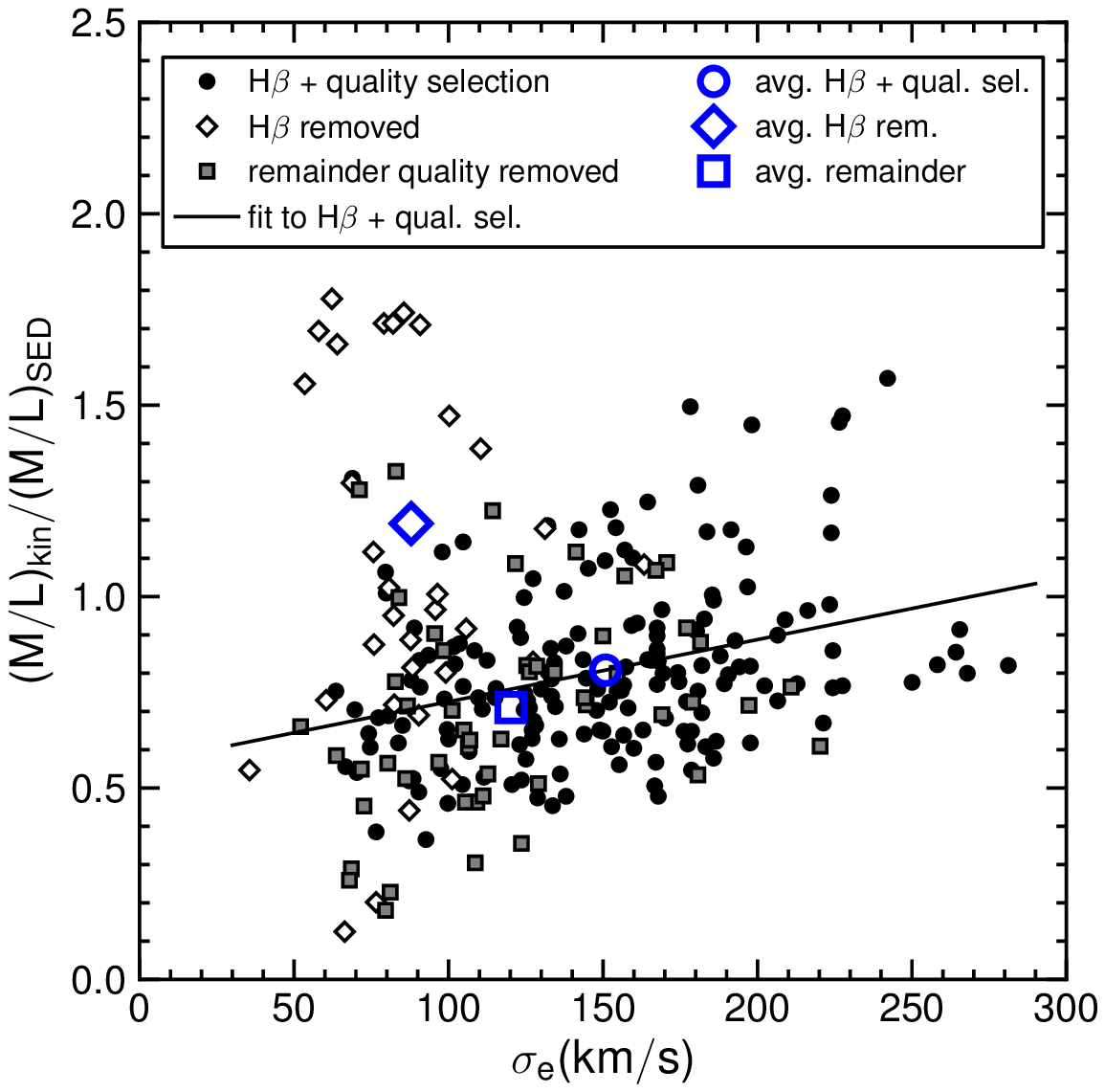}
\includegraphics[width=1.0\columnwidth]{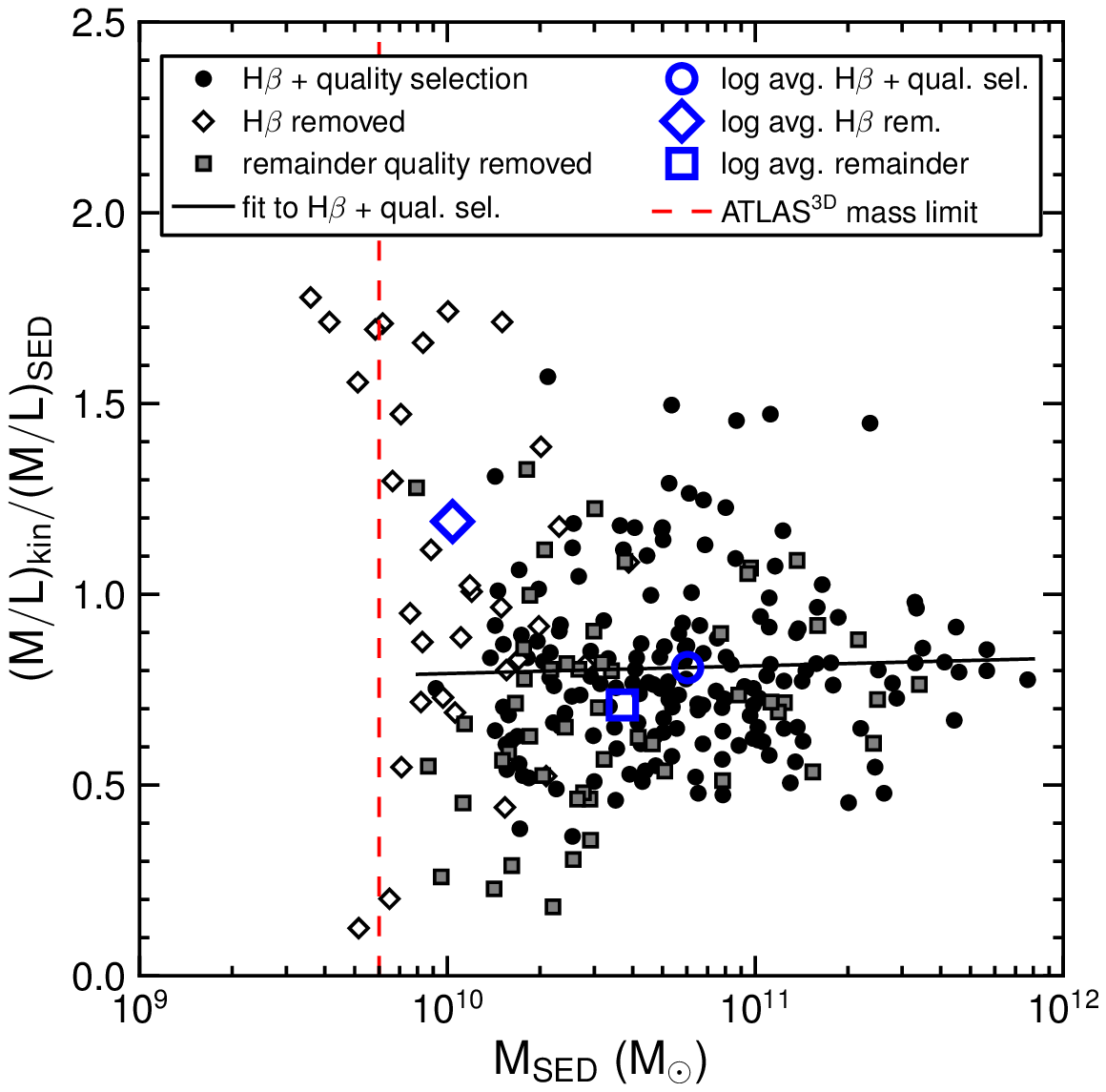}
\caption{ATLAS$^{\rm{3D}}$ data for the IMF mismatch parameter $\alpha=(M/L)_{\rm kin}/(M/L)_{\rm SED}$ as a function of the effective velocity dispersion (left panel) and the mass measured from SED fitting assuming a Salpeter IMF (right panel). The three different data selections are the same as in Figure \ref{GraphStarSalp}. Three open diamond data points with a mismatch parameter greater than 2 are not visible. For each selection the average is indicated in blue. The Pearson $R^2$ coefficients for the fits to the black filled circles are 0.11 for the left panel and 0.001 for the right panel (on a logarithmic mass scale).}
\label{GraphSigmaM}
\end{figure*}

The data show a clear correlation of the mismatch parameter with the effective velocity dispersion $\sigma_e$ \citep{ATLAS15}. Surprisingly, it does not show a correlation with $(M/L)_{\rm SED}$, SDSS colour, luminosity, or even $M_{\rm SED}$ (even though $\sigma_e$ and $M_{\rm SED}$, and $\sigma_e$ and $(M/L)_{\rm SED}$ are tightly correlated). Figure \ref{GraphSigmaM} (left panel) shows the clear trend between the IMF mismatch parameter and the effective velocity dispersion for the high-quality data points (with a Pearson $R^2$ of 0.11).  The variables $\sigma_e$ and $M_{\rm SED}$ are tightly correlated (Pearson $R^2$ of 0.63) so naively one would expect to find a correlation between the IMF mismatch parameter and $M_{\rm SED}$ as well, but Figure \ref{GraphSigmaM} (right panel) shows that this is not the case (Pearson $R^2$ of 0.001). We also see from Figure \ref{GraphSigmaM} (left panel) that the trend with $\sigma_e$ is affected by the exclusion of galaxies with strong H$\beta$ absorption. The excluded galaxies on average have a small $\sigma_e$ and a large $\alpha$. Including all galaxies in the fit of $\alpha$ versus $\sigma_e$ would reduce the best-fit slope from $1.6 \times 10^{-3}$ to $0.4 \times 10^{-3}$ and the Pearson $R^2$ from $0.112$ to $0.003$.

Although the trend of $\alpha$ with $\sigma_e$ is very clear, it is much smaller than the scatter. Accounting for the trend for all galaxies that have not been rejected on basis of H$\beta$ absorption only reduces the scatter in $\alpha$ from 30.6\% to 28.7\%. We note that since velocity dispersion is the main input of the JAM model, we expect it to be more prone to systematics. A very recent analysis of the ATLAS$^{\rm{3D}}$ results by \citet{McDermid14} has found no significant dependence of the IMF on single stellar population equivalent ages or abundance ratios. In the following we will investigate further the IMF mismatch parameter dependence on velocity dispersion, especially in relation to the survey mass completeness and distance measurement effects, but first we take a short look at the implications of the IMF trend on the measurement of galaxy stellar mass functions.

\section{Galaxy Stellar Mass Function and mass completeness}
\label{sectionGSMF}

A non-universal IMF could affect the shape of the galaxy stellar mass function (GSMF) inferred from fitting stellar population synthesis models to the SEDs measured in galaxy surveys.

A recent attempt to quantify this effect is reported by \citet{McGee14}, who take different model assumptions for the dependence of the slope of the IMF on galaxy velocity dispersion and show that the implications for the high-mass end of the GSMF can be quite significant. For such an analysis it makes a difference what observations are taken as the starting point. Also, Figure \ref{GraphSigmaM} suggests that translating an IMF trend with $\sigma_e$ into a trend with $M_{\rm SED}$ can be quite tricky. Here we want to address what would be the effect based solely on the ATLAS$^{\rm{3D}}$ Survey. The advantage of this approach is that we know that the galaxy sample is representative, because it is aimed to be complete down to approximately $6 \times 10^9$ M$_{\odot}$ within the given volume \citep{ATLAS1}.

As can be verified from Figure \ref{GraphSigmaM} (right panel) correcting the observed $M_{\rm SED}$ from any GSMF study to a $M_{\rm kin}$ value, results in the same correction by a factor 0.8 independent of the mass. This just shifts the GSMF of quiescent galaxies to lower masses without changing its shape. Accounting for the scatter in the mismatch parameter (assuming that the scatter is intrinsic and not caused by the observational analysis) would correspond to smoothing the GSMF with a kernel of about 0.2 dex. At the steep high mass end this smoothing kernel effectively shifts the GSMF to lower masses by an additional 0.05 dex. Hence, apart from a possible slight shift of the quiescent GSMF with respect to the star forming GSMF, the ATLAS$^{\rm{3D}}$ results do not imply any changes in the shape of the GSMF.

\begin{figure}
\includegraphics[width=1.0\columnwidth]{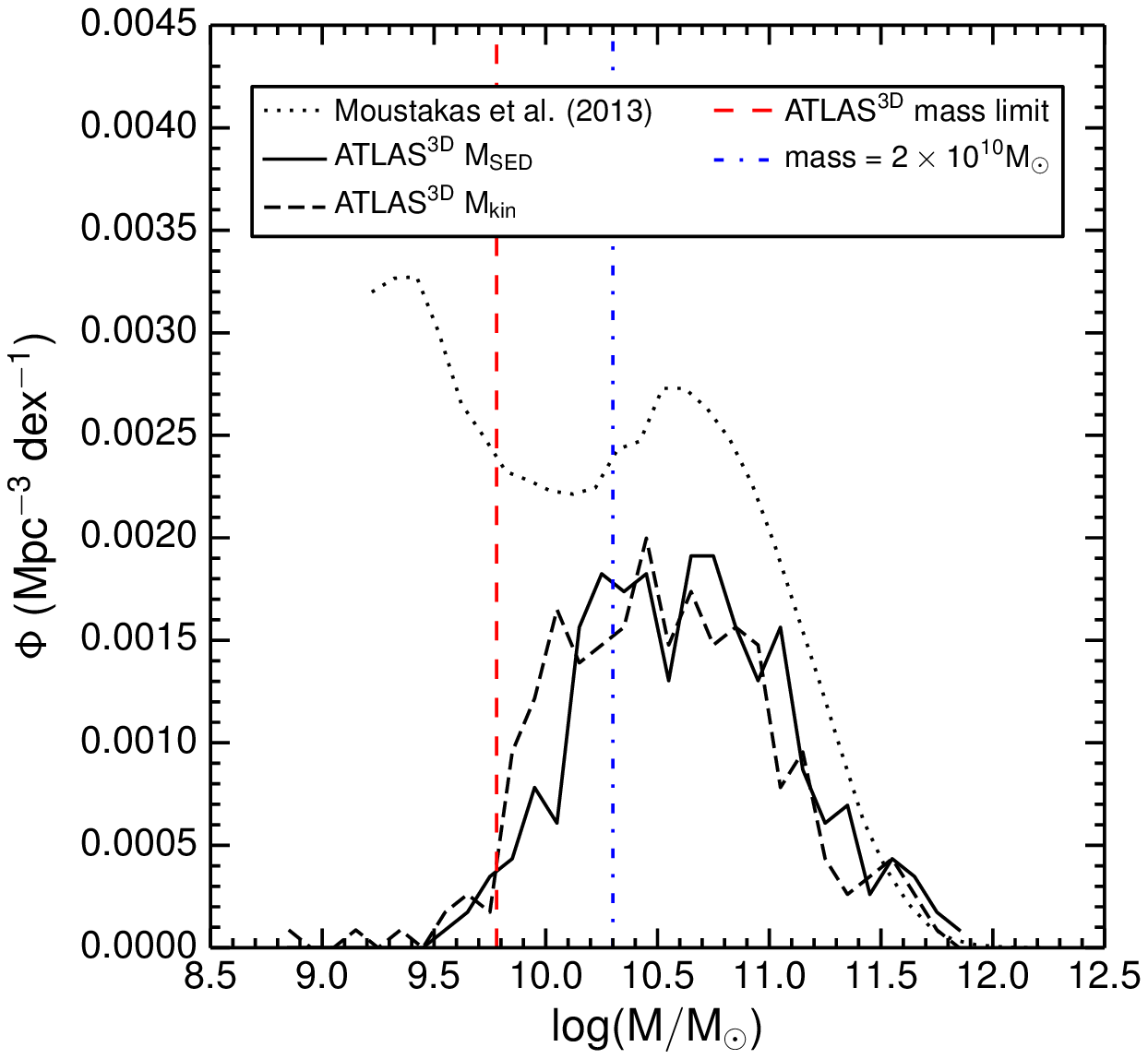}
\caption{The galaxy stellar mass function (GSMF) for all ATLAS$^{\rm{3D}}$ ETGs as a function of respectively $M_{\rm SED}$ from SED fitting (solid curve) or $M_{\rm kin}$ from JAM model fitting (dashed curve).  The JAM model GSMF is shifted to lower masses by a factor of about 0.8, but apart from that there are no major differences. The ATLAS$^{\rm{3D}}$ sample can be compared with the quiescent GSMF from \protect\citet{Moustakas13} (which has been shifted by 0.22 dex in order to correct to a Salpeter IMF). One difference is the overall normalization. On top of that the ATLAS$^{\rm{3D}}$ sample seems to become incomplete already for $M <  2\times10^{10}{\rm M_{\odot} } \approx 10^{10.3}  {\rm M_{\odot} } $ (see Figure \ref{GraphComp} blue dash-dotted line). The high-mass fall off is similar. The red dashed vertical line represents the approximate mass completeness  limit of $6 \times 10^9 {\rm {\rm M_{\odot} }}$ reported by \protect{\citet{ATLAS1}}.}
\label{GraphGSMF}
\end{figure}

We can also look directly at the GSMF for the 260 ETGs in the ATLAS$^{\rm{3D}}$ Survey. For this number of galaxies the statistical and cosmic variation will be quite large, but it is the most direct approach. Figure \ref{GraphGSMF} shows the GSMF separately for $M_{\rm SED}$ and $M_{\rm kin}$ and compares these with the GSMF for quiescent galaxies from \citet{Moustakas13}. Apart from the overall shift in mass by a factor 0.81, the two ATLAS$^{\rm{3D}}$ mass determinations give very similar GSMFs. The high-mass fall off from ATLAS$^{\rm{3D}}$ is the same as that from \citet{Moustakas13}. The overall normalization is approximately 30\% lower, which could be due to cosmic variance or a difference in selection criteria for quiescence. At the low-mass end the ATLAS$^{\rm{3D}}$ GSMF falls off rapidly, which most likely indicates that the galaxy sample is incomplete\footnote{Alternatively this could be caused by a divergence of the selection criteria on quiescence from \citet{Moustakas13} with respect to the ETG sample of ATLAS$^{\rm 3D}$, which occurs abruptly at masses $M \lesssim 10^{10.3} {\rm M_{\odot} }$.}. 

\begin{figure*}
\includegraphics[width=1.0\columnwidth]{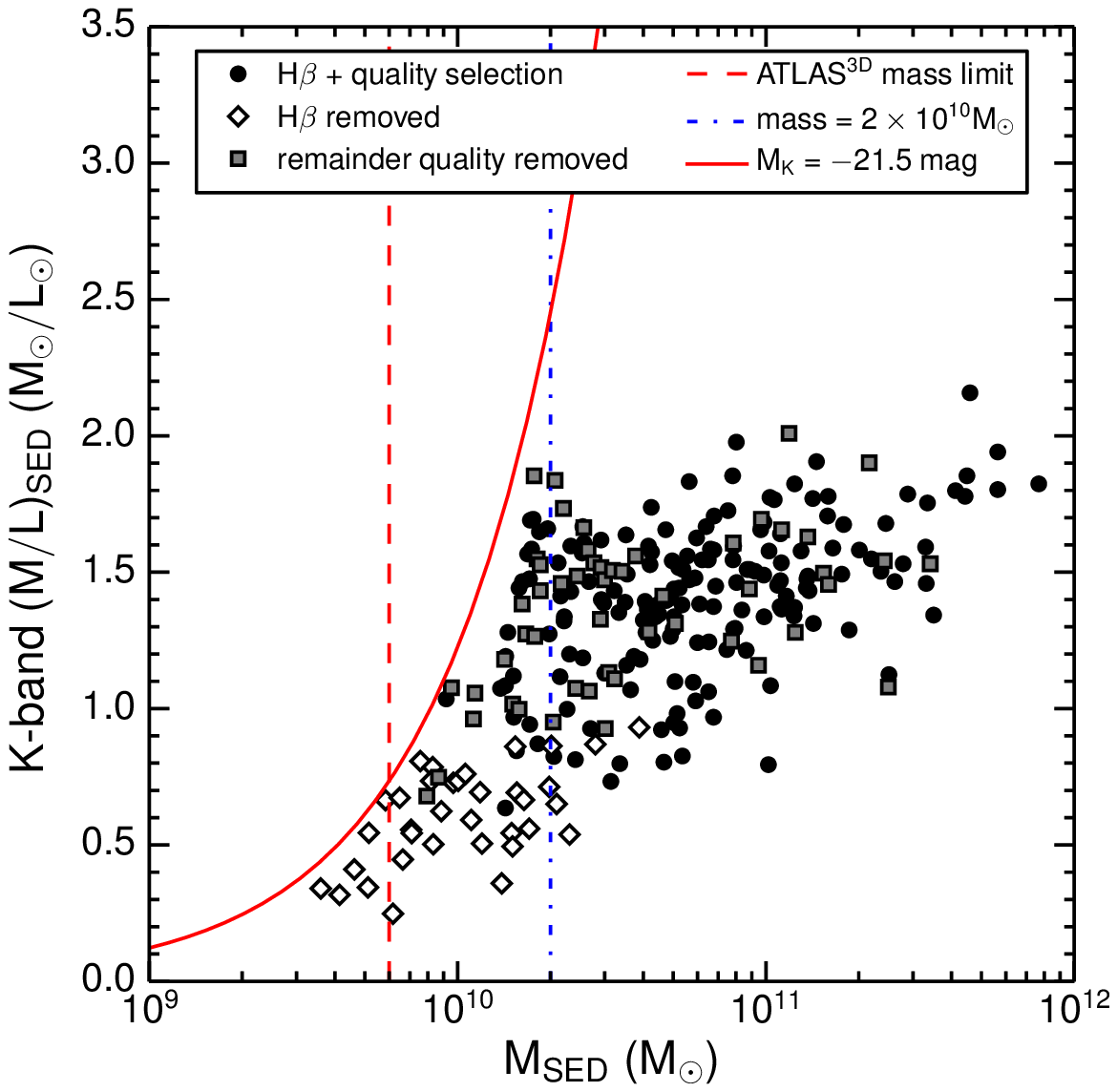} 
\includegraphics[width=1.0\columnwidth]{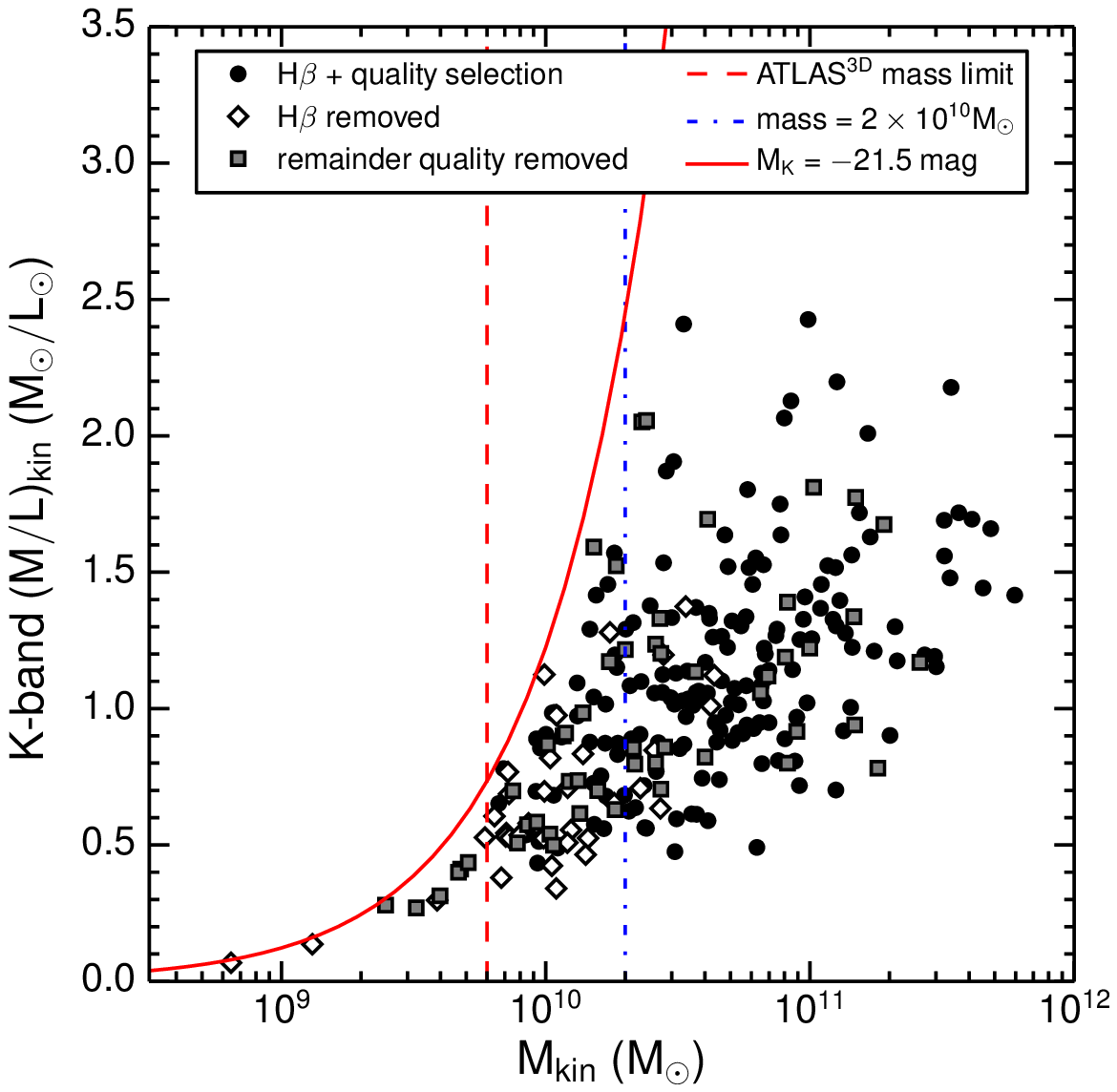}
\includegraphics[width=1.0\columnwidth]{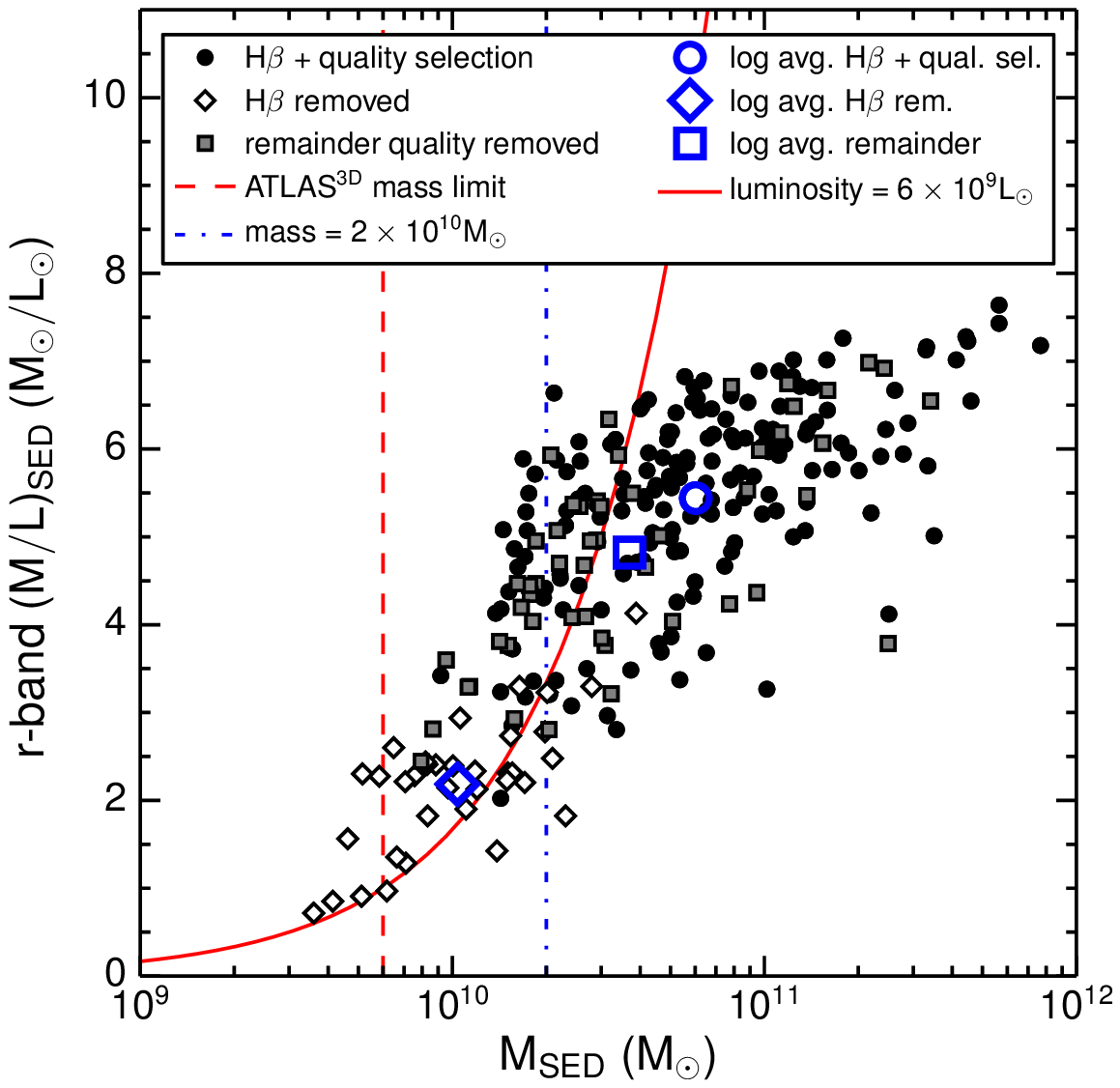}
\includegraphics[width=1.0\columnwidth]{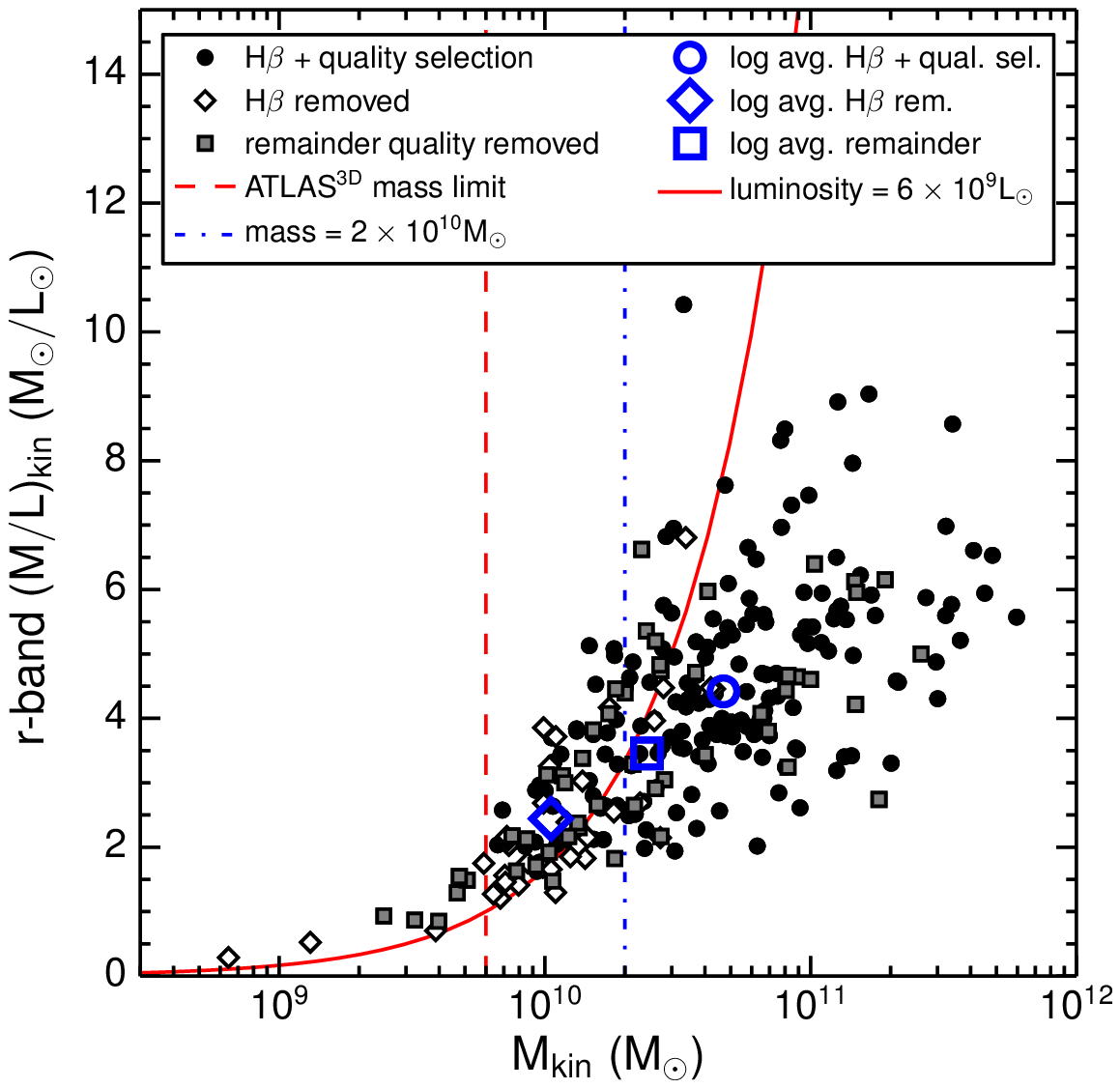}
\caption{$(M/L)$ versus $M$ for both mass determinations. The data sets are the same as in previous figures. The left panels indicate the mass from SED fitting, the right panels indicate the kinetic mass determination by ATLAS$^{\rm{3D}}$. The top row corresponds to K-band luminosities, with the solid red line indicating the selection limit of $M_{K} = -21.5$ $mag$. The bottom row corresponds to the r-band luminosities used throughout the rest of this article and in the definition of the IMF mismatch parameter $\alpha$. The upper row clearly demonstrates that the selection is not complete for masses (either $M_{\rm{SED}}$ or $M_{\rm{kin}}$) below $2 \times 10^{10} \rm{M_{\odot}}$. Under this limit galaxies with high K-band (M/L) are not selected. The blue dash-dotted line represents this conservative mass limit. The red dashed vertical lines denotes the approximate ATLAS$^{\rm{3D}}$ survey limit of $6 \times 10^9 {\rm {\rm M_{\odot} }}$ reported by \protect\citet{ATLAS1}. The lower row shows the same completeness behavior in the r-band. For reference the red solid curve in the bottom panels indicates a constant luminosity of $6 \times 10^9 L_{\odot}$. Also in the r-band the selection edge runs roughly parallel to this constant luminosity curve. }
\label{GraphComp}
\end{figure*}

In order to assess the mass completeness of the survey, Figure \ref{GraphComp} (top row) shows the mass to light ratio in the K-band as a function of $M_{\rm{SED}}$ (left) and $M_{\rm{kin}}$ (right). The galaxy selection is based on K-band luminosity ($M_K< -21.5$ $mag$). The selection is made in the K-band because $(M/L)$ variations in the K-band are smaller than in the r-band. For masses smaller than $2\times10^{10} {\rm M_{\odot}}$ the survey is not mass complete. Galaxies with masses below this limit are bound by a progressively smaller upper limit on the K-band $(M/L)$. Figure \ref{GraphComp} (bottom row) shows that the r-band $(M/L)$ follows the same mass completeness trend. The mass-completeness limit of  $2\times10^{10} {\rm M_{\odot}}$ that we estimate is higher than the survey limit of $M \approx 6 \times 10^{9} {\rm M_{\odot} }$ reported by \citet{ATLAS1}.

One should be cautious about the biases that these completeness effects might introduce. For instance, the Mass Plane projection of $\sigma$ versus $M$ of \citet{ATLAS20} has selected against red, high $M/L$ galaxies with masses below $2\times10^{10} {\rm M_{\odot}}$. Inclusion of such galaxies might change the $M/L$ dependence on $M$ and $\sigma$ significantly at the low mass end.

\section{Mass completeness effects on the IMF dispersion trend}
\label{SectionCompleteness}

\begin{figure}
\includegraphics[width=1.0\columnwidth]{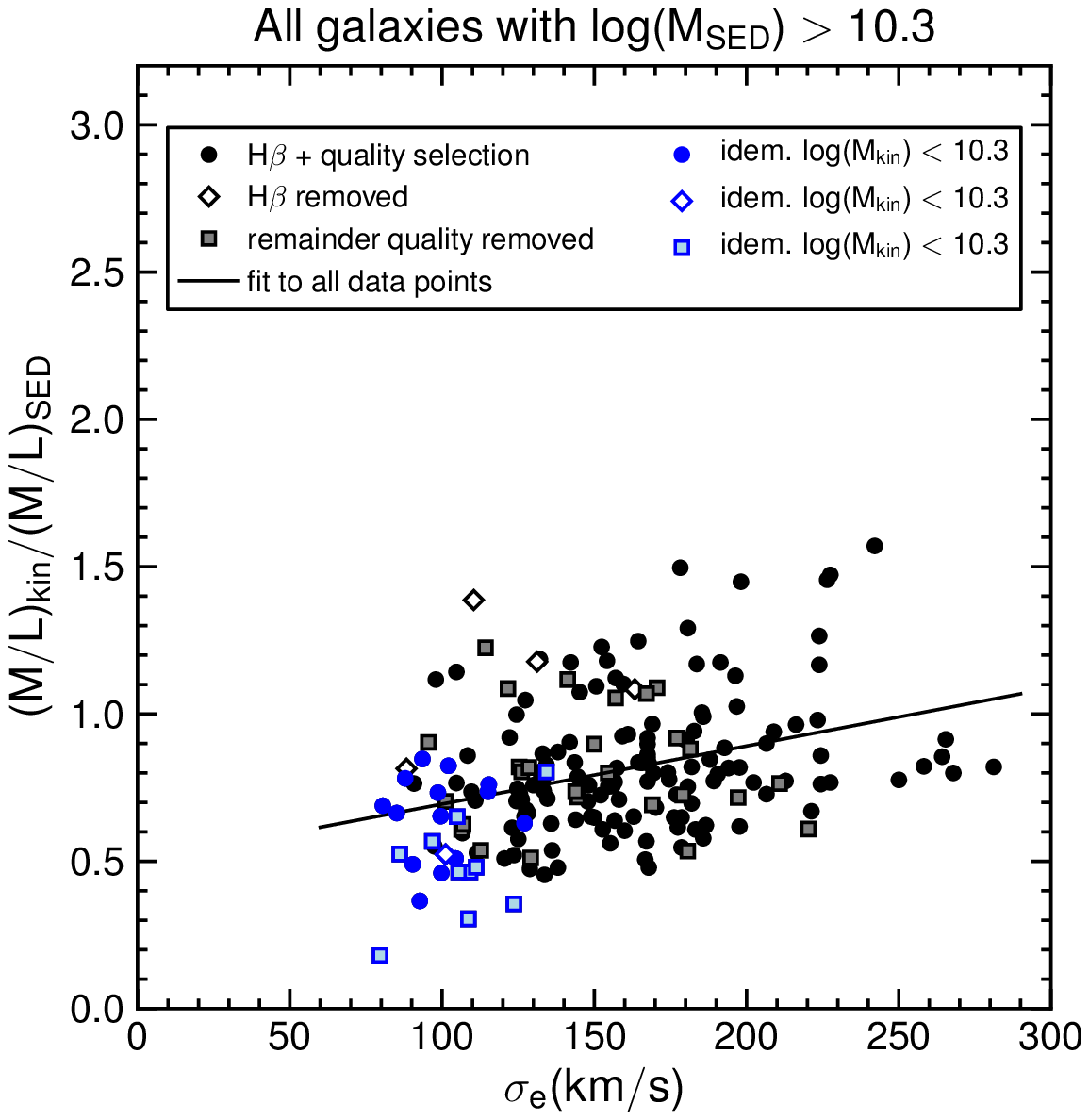}
\includegraphics[width=1.0\columnwidth]{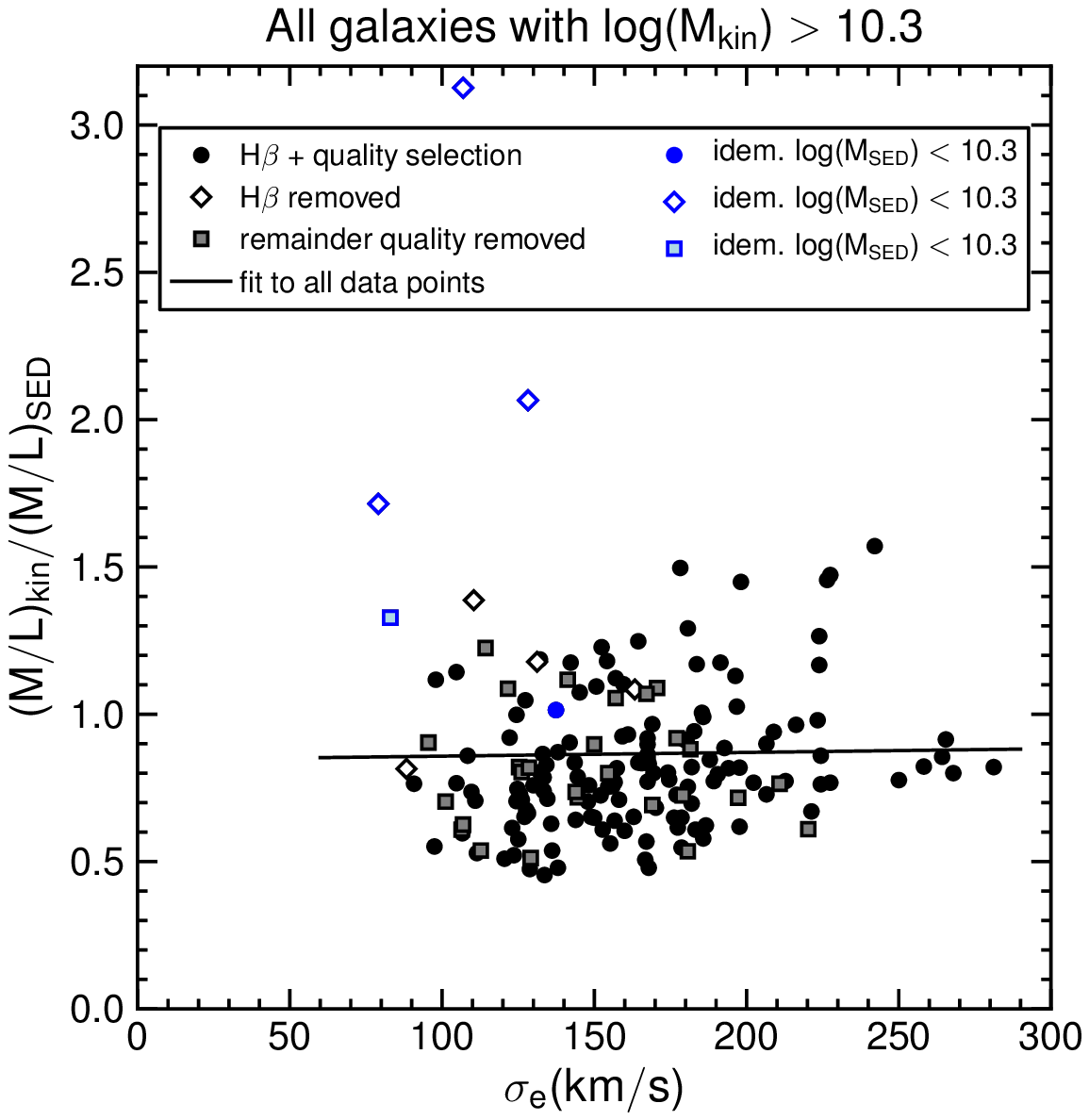}
\caption{Trend of the IMF mismatch parameter $\alpha=(M/L)_{\rm kin}/(M/L)_{\rm SED}$ with velocity dispersion for all ETGs with $M_{\rm SED}>10^{10.3}{\rm {\rm M_{\odot} }}$ (top panel), or $M_{\rm kin}>10^{10.3}{\rm {\rm M_{\odot} }}$ (bottom panel). the top sample gives a slope of 0.0020, Pearson $R^2 = 0.12$, Spearman $R^2 = 0.13$. The bottom sample gives a slope of 0.0001, Pearson $R^2 = 0.0003$, Spearman $R^2 = 0.02$. Data points in blue indicate galaxies that are only present in one of the two panels.}
\label{GraphMassConstraint}
\end{figure}

In the previous section we showed that the ATLAS$^{\rm{3D}}$ Survey is probably incomplete for galaxy masses below $10^{10.3} {\rm {\rm M_{\odot} }}$. This introduces a complex bias. Most of the problematic galaxies, especially those with non-homogenous $(M/L)$ ratios caused by recent star formation, also have masses below this limit. It therefore makes sense to look at the mass-complete sample of galaxies with masses higher than $10^{10.3} {\rm {\rm M_{\odot} }}$. There are two possible ways to implement this. We can either impose a cut in $M_{\rm SED}$ or in $M_{\rm kin}$. Figure \ref{GraphMassConstraint} shows the IMF trends obtained by imposing either of these constraints. Using a $M_{\rm SED}$ cut gives a very clear  IMF trend with $\sigma_e$, whereas using a $M_{\rm kin}$ cut gives no trend at all. It is straightforward to understand what is the cause of this difference. Around galaxy masses of $10^{10.3} {\rm {\rm M_{\odot} }}$ the first selection will favour high $(M/L)_{\rm SED}$ galaxies and hence low $\alpha$, while the second selection will favour high $(M/L)_{\rm kin}$ galaxies and hence high $\alpha$. The region where this selection effect shows up in a $(\sigma_e,\alpha)$ plot is at low $\sigma_e$, because of the tight correlation between velocity dispersion and mass.

It seems that the IMF trend with velocity dispersion depends on the mass selection criterion. For a mass complete sample of galaxies with $M_{\rm{kin}} > 2 \times 10^{10} {\rm M_{\odot} }$ one would conclude that there is no IMF trend with velocity dispersion over a large range of velocities. At the moment, the conclusions that we draw about the IMF dependency on velocity dispersion are dominated by the precise selection criterion at the low mass or low luminosity end of the galaxy sample. Therefore in the future it would be very useful to push this limit towards lower masses and lower luminosities.

\section{Distance effects and SBF calibration}
\label{SectionDistance}

A source of error or bias in the determination of the IMF mismatch parameter lies in the distance determination. ATLAS$^{\rm{3D}}$ looks at nearby galaxies. For these galaxies the relative error in redshift-distances can be large. The distances used in the JAM method come from various sources: SBF distances from \citet{Tonry01} and \citet{Mei07}, distances from the NED-D Catalogue and distances from the redshift, via the local flow field model of \citet{Mould00} (using only the Virgo attractor).

The inferred value of $(M/L)_{\rm SED}$ is independent of the distance determination, but $(M/L)_{\rm kin}$ does depend on the distance. Suppose that the distance is overestimated by a factor $\eta$. This would mean that the luminosity of the galaxy is overestimated by a factor $\eta^2$ and that the size of the galaxy is overestimated by a factor $\eta$. Since the JAM fitting method is in effect a sophisticated way of determining a dynamical mass, the mass will follow $M \propto \sigma^2 r$ and will be overestimated by a factor $\eta$. This means that $(M/L)_{\rm kin}$ and hence the IMF mismatch parameter, will be a factor $\eta$ too small. Thus, if a galaxy in reality is closer than determined, it will have a higher $(M/L)_{\rm kin}$ than determined and vice versa. Any errors and biases in the distance determination will therefore show up as errors and biases in the IMF determination\footnote{For galaxies around the completeness limit of sections \ref{sectionGSMF}, \ref{SectionCompleteness} the selection on K-band intrinsic luminosity will contain some galaxies that should fail the selection criterium, but are included due to an overestimate of the distance. This distance error propagates quadratically into the intrinsic luminosity. For these galaxies the perceived IMF mismatch parameter will be too small. Vice versa some galaxies with underestimated distances will be missed.}.

\begin{figure}
\includegraphics[width=1.0\columnwidth]{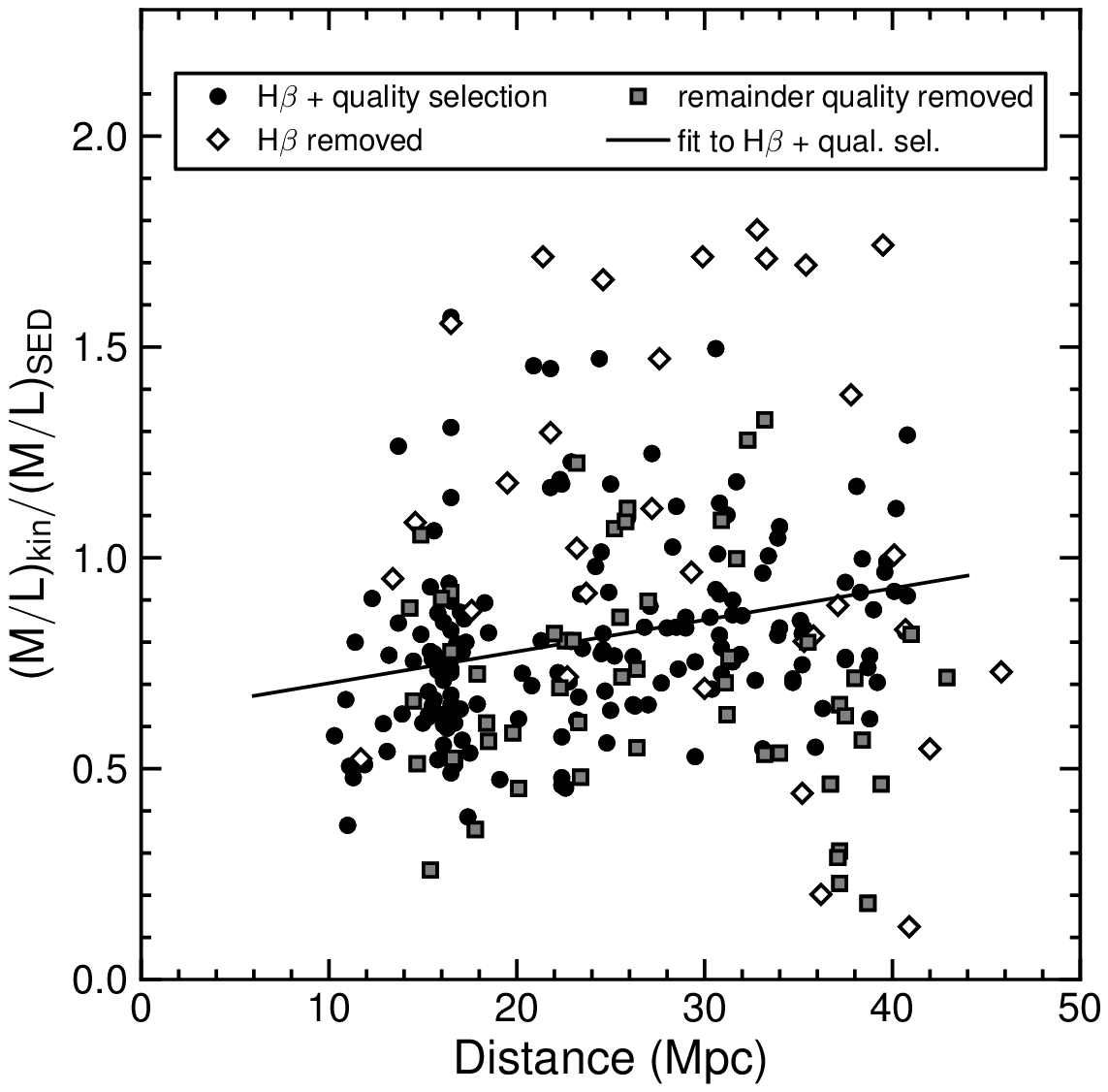}
\includegraphics[width=1.0\columnwidth]{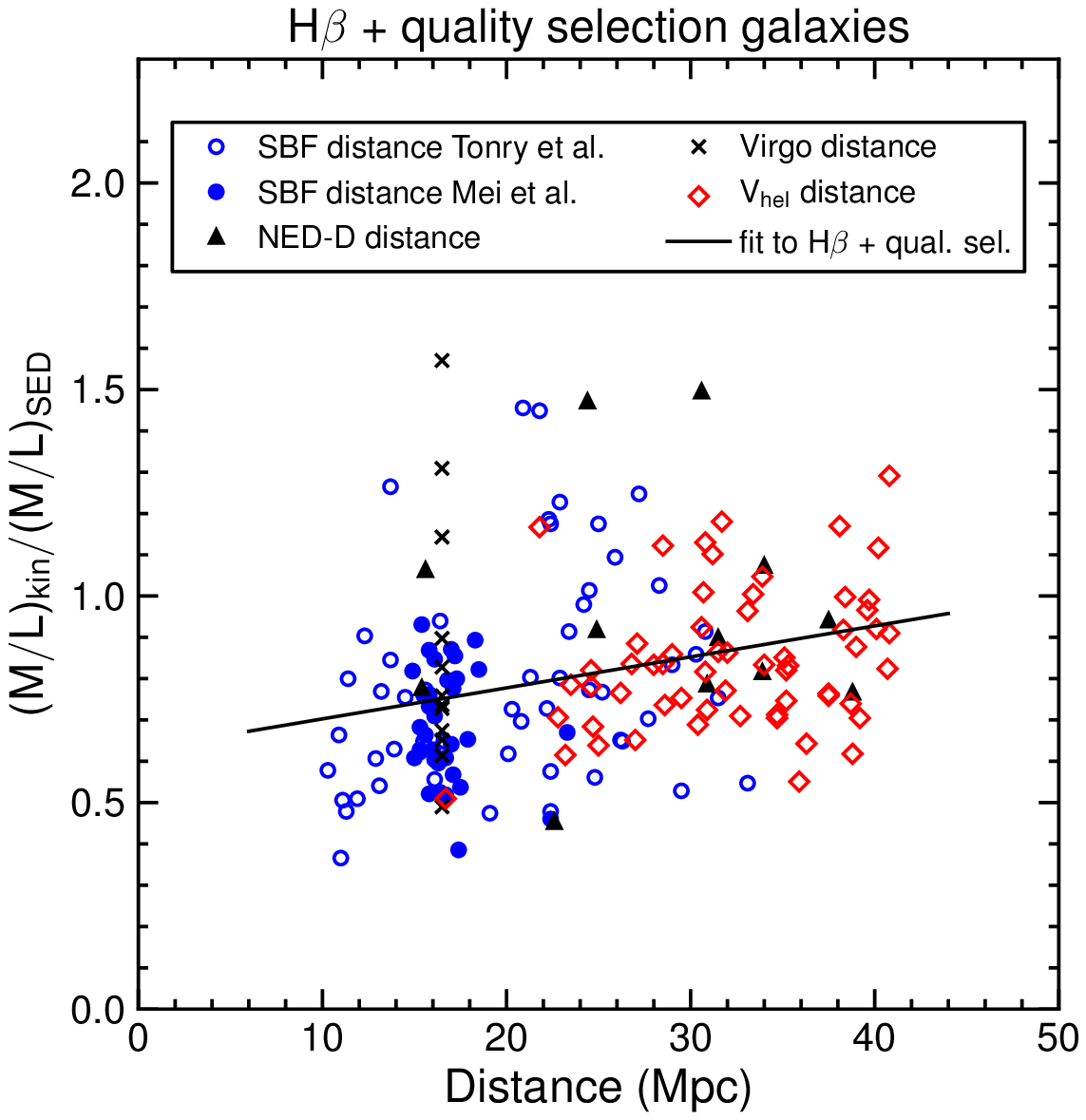}
\caption{The IMF mismatch parameter $\alpha=(M/L)_{\rm kin}/(M/L)_{\rm SED}$ versus distance. Top panel: for the high-quality galaxies there is a clear trend with distance (Pearson $R^2=0.08$, Spearman $R^2=0.12$). Three open diamonds with a mismatch parameter larger than 2 are situated beyond the plotted range in the upper right. Bottom panel: only the galaxies selected on quality and H$\beta$ absorption. The different symbols indicate the source of the distance measure that is used as input in the JAM fitting procedure. Blue open circles correspond to distances from \protect{\citet{Tonry01}}, blue filled circles correspond to distances from \protect{\citet{Mei07}}, black triangles are from the NED-D catalogue, black crosses indicate galaxies for which the distance is set at the distance of the Virgo cluster, red open diamonds correspond to distances via the heliocentric redshift velocity. The solid line in both panels is the same fit to the high quality galaxies.}
\label{GraphDistance}
\end{figure}

Figure \ref{GraphDistance} (top panel) shows the dependence of the IMF mismatch parameter on distance. For the high-quality galaxies there is a trend of increasing IMF mismatch parameter with distance. One possibility is that this reflects a genuine systematic variation in the IMF on Mpc scales. If this were due to a dependence of the IMF on environment, then one would expect a stronger correlation between the IMF mismatch parameter of neighbouring galaxies. For example, \citet{ATLAS7} have used the ATLAS$^{\rm{3D}}$ data to show that the morphology of the galaxies depends on their immediate environment (the galaxy density defined by the closest three galaxies). However, we find no appreciable correlation between the IMF mismatch parameter of nearest neighbours  (Pearson $R^2=0.02$, Spearman $R^2=0.03$).

Another possibility would be that the distance trend of the IMF mismatch parameter is related to the mass completeness issues from the previous section. This could be the case if the survey would have missed galaxies with low masses at larger distances. This is however not the case. There is no trend with distance for either velocity dispersion, kinematic mass or SED fitting mass (respective Pearson $R^2$ of 0.0008, 0.0005 and 0.0005).

This leaves the possibility that the IMF trend with distance is possibly caused by a bias in the distance determination. Figure \ref{GraphDistance} (bottom panel) shows the different sources for the distances that are used as input in the JAM fitting method.  A relative distance error eventually translates into a relative error in the IMF mismatch parameter. Table \ref{TableStats} gives the mean and standard deviation of the IMF mismatch parameter for each set of distances. The ratio $\sigma(\alpha)/\overline{\alpha}$ is smallest for the samples using the SBF distances from \citet{Mei07} and the distances from the redshift via the local flow field model of \citet{Mould00}, suggesting that these methods give the highest relative accuracy. The other three sets are considerably worse.

There is no clear cut way to unambiguously prove which distance method is causing the bias. Part of the overall correlation between $\alpha$ and distance is caused by the offset of the SBF distance determinations at small distances with the redshift distance determination at larger distances and part of it is caused by correlations within each data set. These correlations within each data set are biased by the selection effect of which galaxy belongs to which data set. Especially in the region around 25 Mpc, the choice between ``Tonry'' and ``$\rm{V_{hel}}$'' can itself cause a correlation between the IMF mismatch parameter and distance of the corresponding subsets of galaxies. It is therefore better to look at a selection criterion based on distance ($D<25$ {Mpc} versus $D>25$ {Mpc}), which overlaps with the regions where both distance methods are used. Table \ref{TableCorrelation} shows that the trend of the IMF mismatch parameter with distance originates from the galaxies closer than 25 Mpc that are not a member of the Virgo Cluster. This might point towards a bias in the SBF distance determination from \citet{Tonry01}.

\begin{table}
\begin{center}
\begin{tabular}{lllll}
\hline
distance method & $\overline{\alpha}$ & $\sigma(\alpha)$ & $\sigma(\alpha)/\overline{\alpha}$ \\   \hline
SBF Mei & 0.69 & 0.14 & 0.20 \\ 
SBF Tonry & 0.79 & 0.26 & 0.33 \\ 
NED-D &  0.96 & 0.30 & 0.31 \\ 
Virgo & 0.89 & 0.32 & 0.36 \\ 
$\rm{V_{hel}}$ & 0.85 & 0.17 & 0.20 \\  \hline
\end{tabular}
\caption{Average IMF mismatch parameter $\overline{\alpha}$ and the standard deviation $\sigma(\alpha)$ for the galaxy samples corresponding to different methods to measure their distances: ``SBF Mei'' refers to galaxies with a distance determination by \citet{Mei07}, ``SBF Tonry'' refers to distances by \citet{Tonry01}, ``NED-D'' are galaxies for which the distance is taken as the average of NED-D catalogue values, ``Virgo'' are galaxies whose distance is set equal to the distance of the Virgo cluster, ``$\rm{V_{hel}}$'' are galaxies for which the distance is determined from their heliocentric redshift velocity.}
\label{TableStats}
\end{center}
\end{table}

\begin{table*}
\begin{center}
\begin{tabular}{llllllll}
\hline
 & & &($\alpha$ vs. D)  & & & ($\alpha$ vs. $\sigma_e$) & \\
Galaxy Selection & number of galaxies & Pearson $R^2$ & p-value & slope & Pearson $R^2$  & p-value & slope  \\   
&  & Spearman $R^2$ & p-value &  &  Spearman $R^2$  & p-value &   \\   \hline
H$\beta$ + quality selection & 171 & 0.08 & 0.0002 & \hspace{2.7pt}0.0075  & 0.11 & 0.000008 & 0.0016 \\ 
 & & 0.12 & 0.000003 &  & 0.12 & 0.000005 & \\  
D $>$ 25 Mpc & 70 &0.006 & 0.54 & \hspace{2.7pt}0.0031  & 0.05 & 0.05 & 0.0009  \\   
& & 0.008 & 0.46 & &  0.08 & 0.02 & \\  
Virgo & 47 &  0.0005 & 0.88 & -0.0062  & 0.05 & 0.13 & 0.0009   \\   
&  &0.001 & 0.82 &  & 0.06 & 0.10 & \\  
D $<$ 25 Mpc, non-Virgo & 52 &  0.11 & 0.01 & \hspace{2.7pt}0.0185  & 0.29 & 0.00003 & 0.0033 \\   
& & 0.12 & 0.01 &  & 0.30 & 0.00003 & \\  
$\rm{V_{hel}}$ &  59 &0.07 & 0.04 & \hspace{2.7pt}0.0080 &  0.12 & 0.006 & 0.0014  \\   
&  &0.06 & 0.06 &  & 0.14 & 0.004 & \\  
SBF Mei & 34 &0.04 & 0.28 & -0.0145  & 0.17 & 0.01 & 0.0010   \\  
&  &0.01 & 0.57 &  & 0.16& 0.02 & \\  
SBF Tonry & 54 & 0.08 & 0.04 & \hspace{2.7pt}0.0112 &  0.21 & 0.0004 & 0.0027  \\ 
&  &0.11 & 0.02 & &  0.27 & 0.00007 & \\  \hline
\end{tabular}
\caption{Trend of the IMF mismatch parameter with both distance $D$ (columns 3-5) and effective velocity dispersion $\sigma_e$ (columns 6-8). Both trends are quantified by the Pearson $R^2$ coefficient, by the corresponding 2-tailed p-value for the null-hypothesis of no correlation and by the slope of the best linear fit. The Spearman $R^2$ and corresponding 2-tailed p-value are also given. Note that at a fixed slope, $R^2$ increases if the scatter decreases, thus ``SBF Mei'' and ``$\rm{V_{hel}}$'' naturally have a higher $R^2$ coefficient. The first row corresponds to all high-quality galaxies, selected by having a non-zero ``quality'' label in \citet{ATLAS15} and an H$\beta$ absorption with an equivalent width smaller than $2.3$ \AA. The next three rows are subsets of these high quality galaxies based on distance, where the galaxies with distances smaller than 25 Mpc have been split into Virgo galaxies and non-Virgo galaxies.  The last three rows correspond to subsets defined by different distance determination methods:  ``$\rm{V_{hel}}$'' are galaxies for which the distance is determined from their heliocentric redshift velocity, ``SBF Mei'' refers to galaxies with a distance determination by \citet{Mei07}, ``SBF Tonry'' refers to distances by \citet{Tonry01}. The other two distance methods from Table \ref{TableStats} are not included, because both contain only 12 galaxies, too few to give meaningful statistics.  Both the trend with $D$ and the trend with $\sigma_e$ are mostly due to the non-Virgo, $D<25$ Mpc set or, equivalently, the sample of galaxies with a distance determination from \protect\citet{Tonry01}. The trends with $\sigma_e$ are also plotted in Figure \ref{GraphABCDEF}.}
\label{TableCorrelation}
\end{center}
\end{table*}

\begin{figure}
\includegraphics[width=0.49\columnwidth]{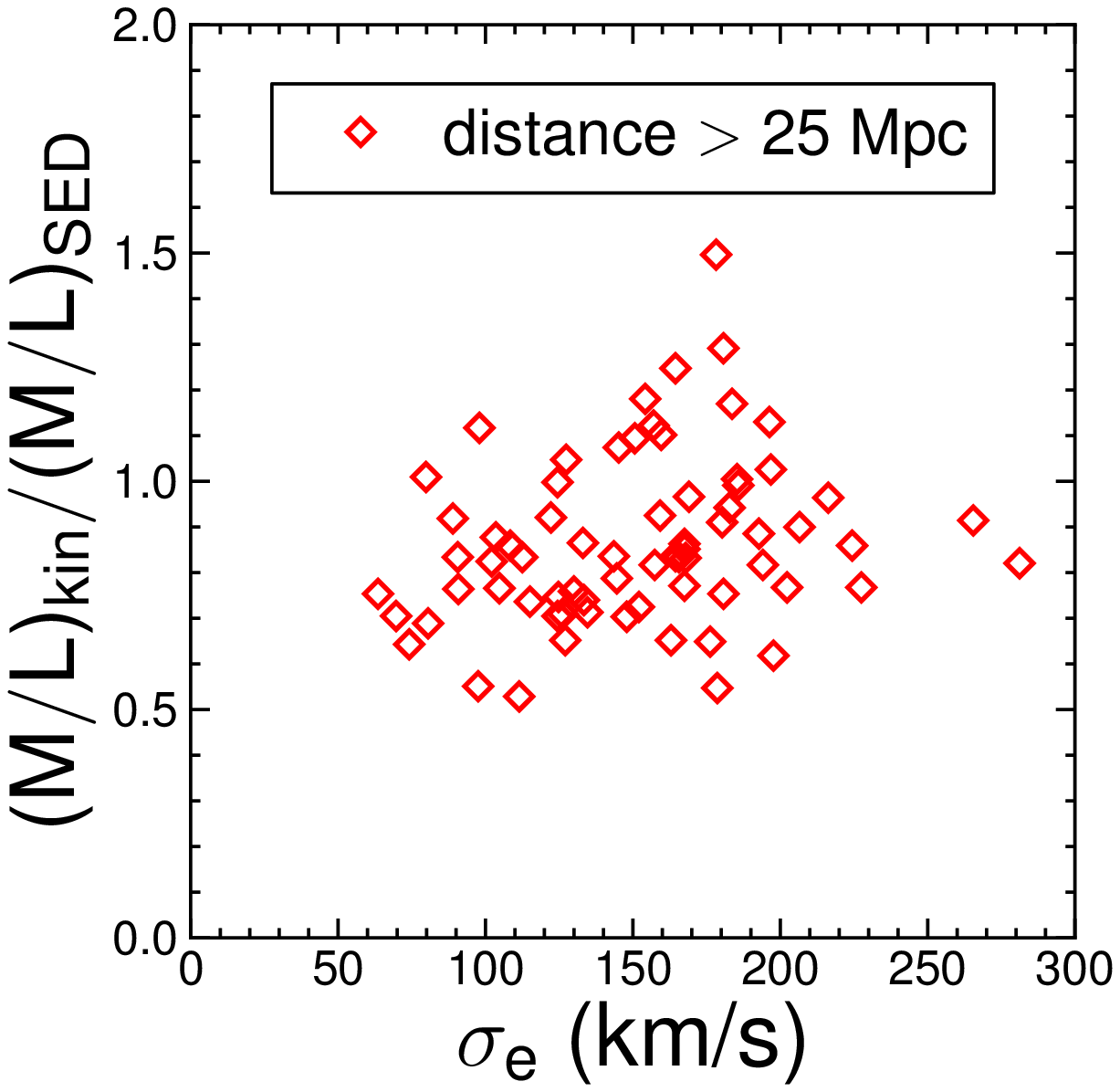}
\includegraphics[width=0.49\columnwidth]{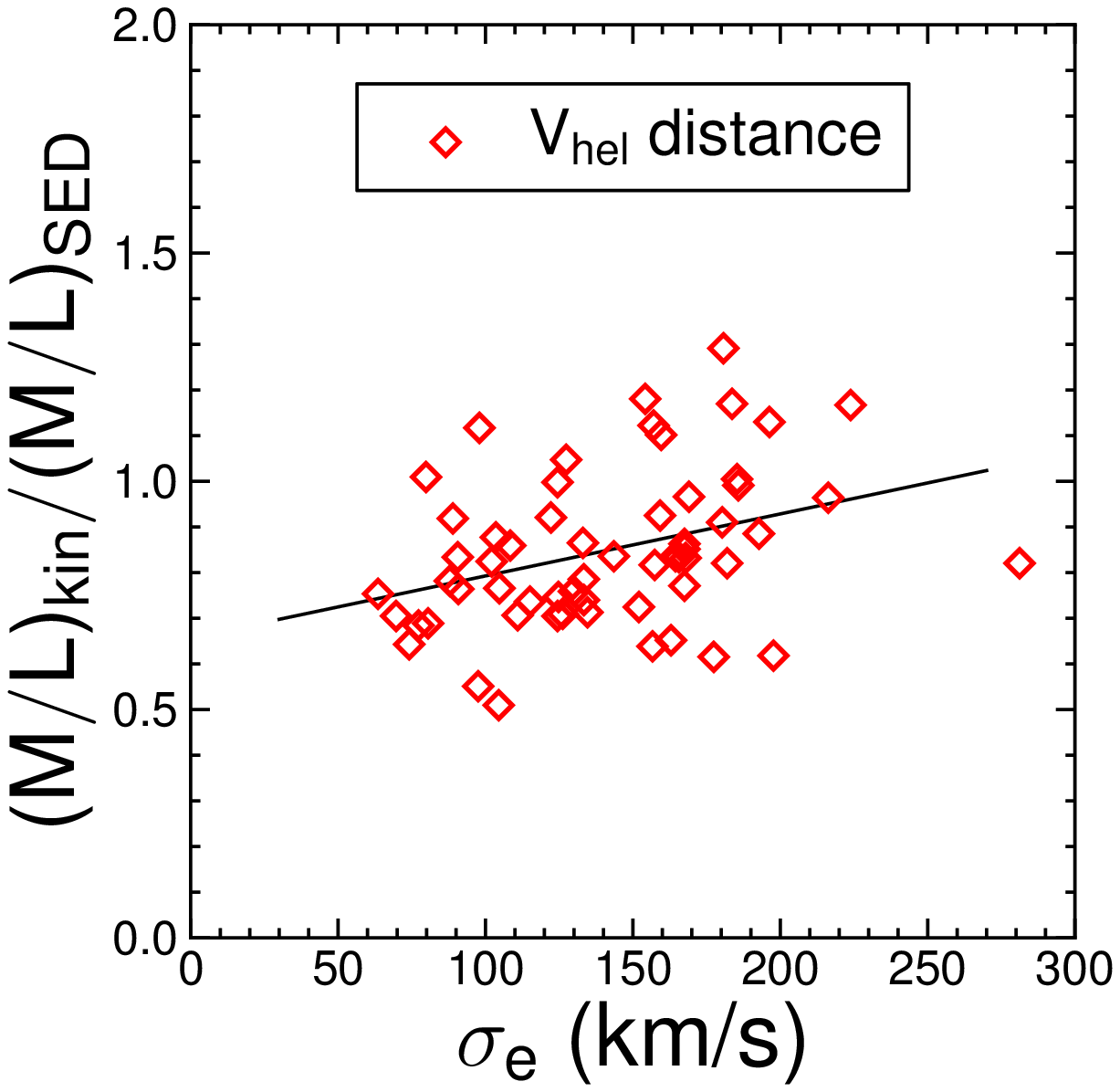}\\
\includegraphics[width=0.49\columnwidth]{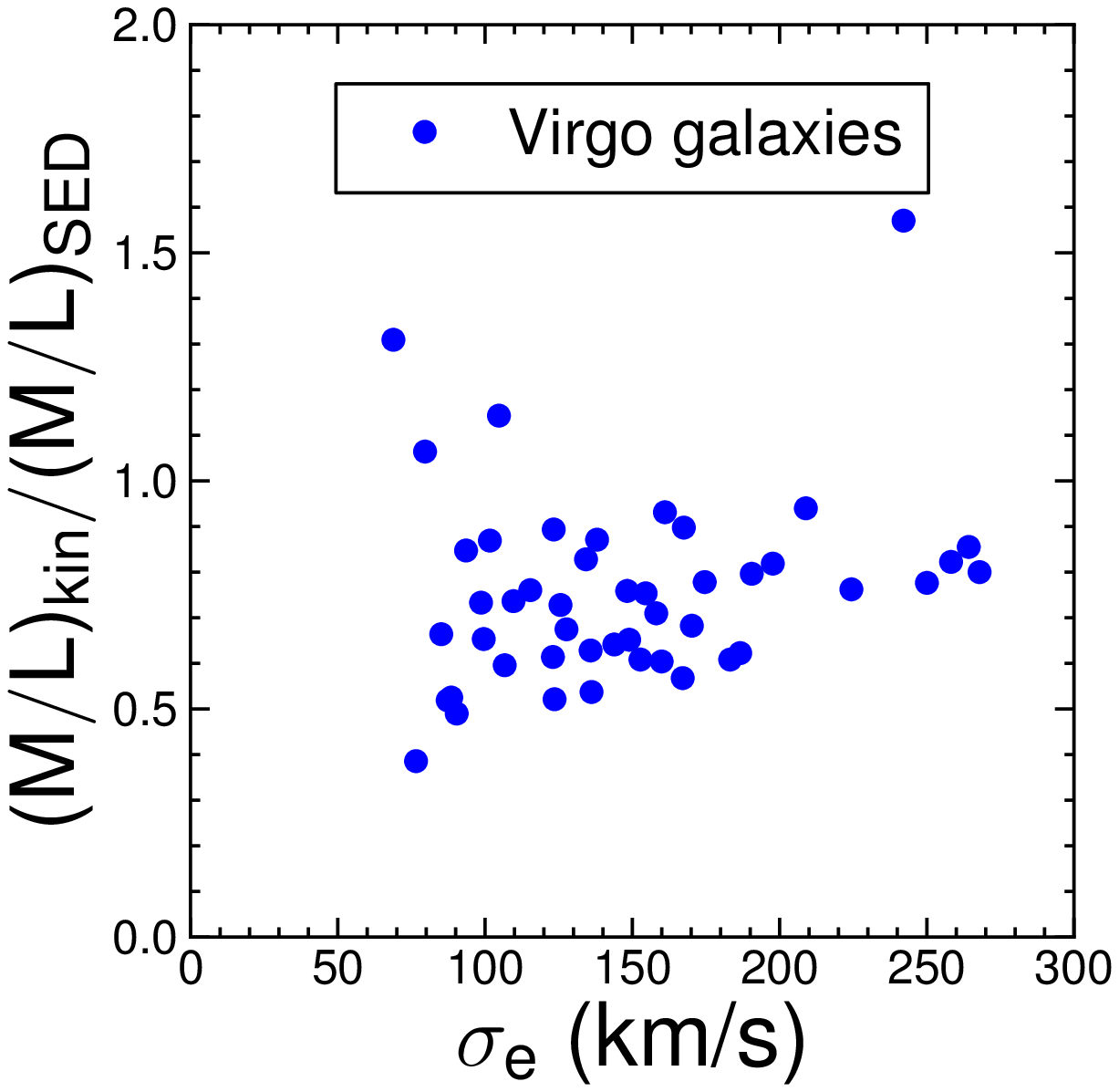}
\includegraphics[width=0.49\columnwidth]{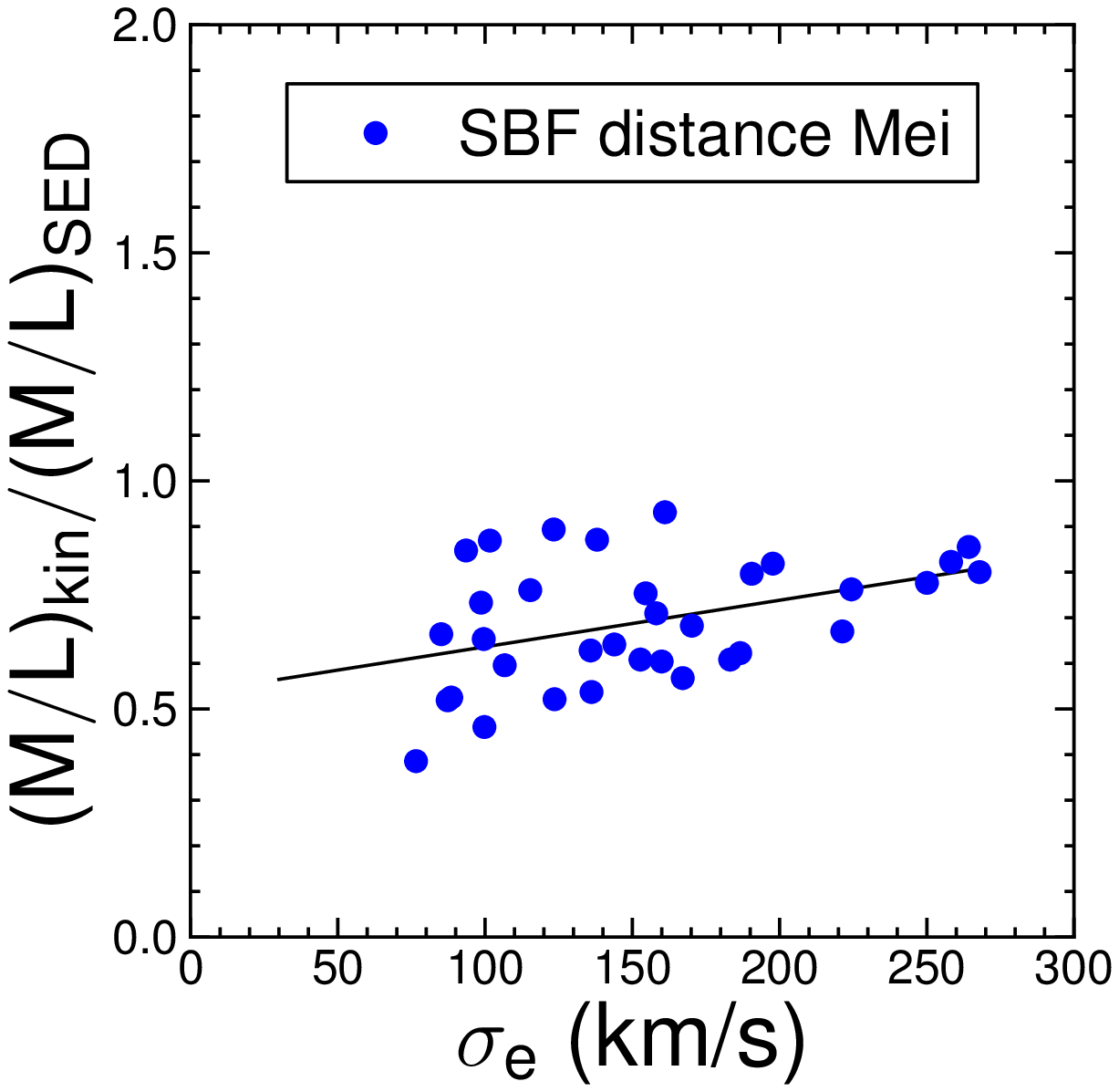}\\
\includegraphics[width=0.49\columnwidth]{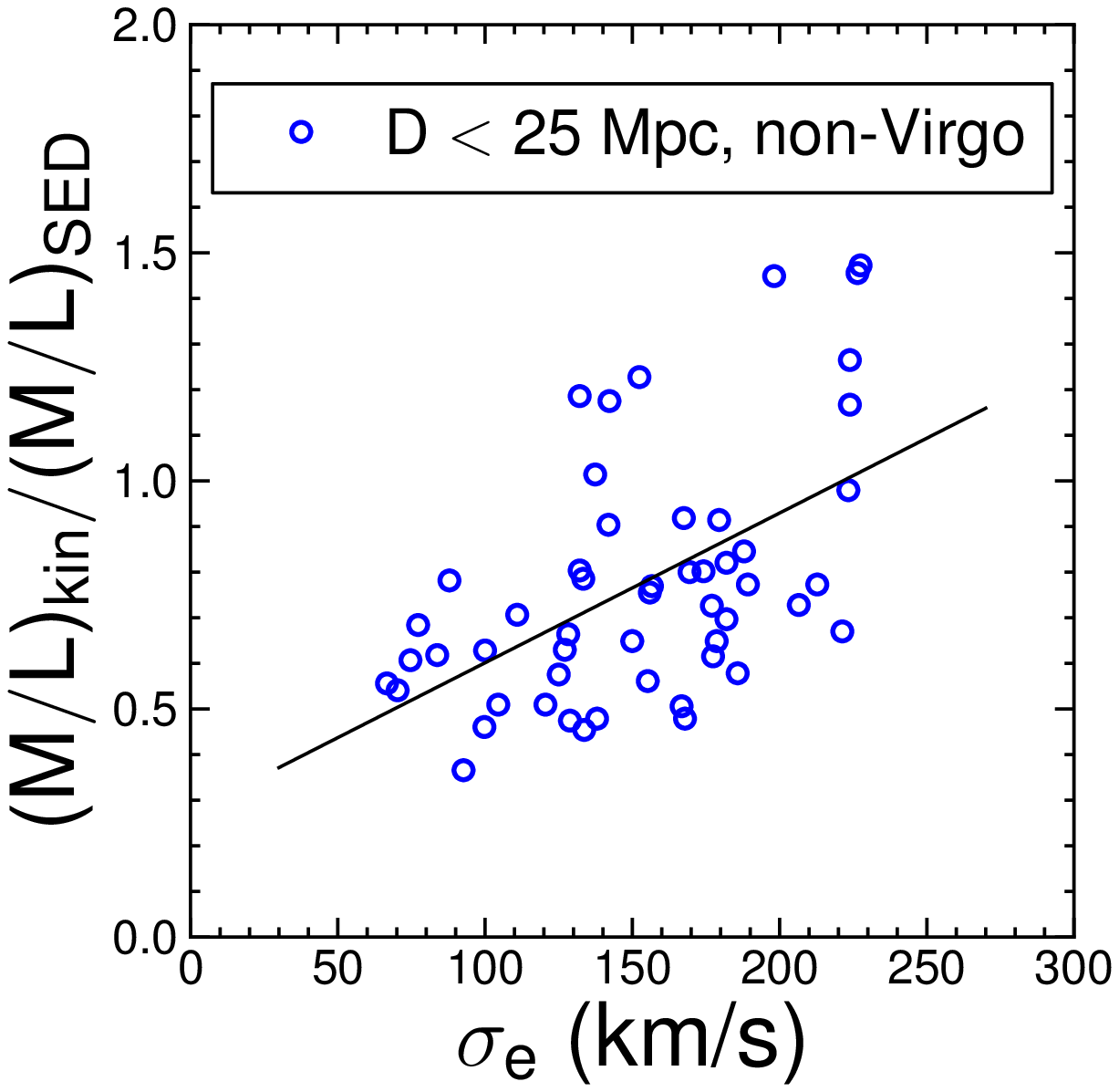} 
\includegraphics[width=0.49\columnwidth]{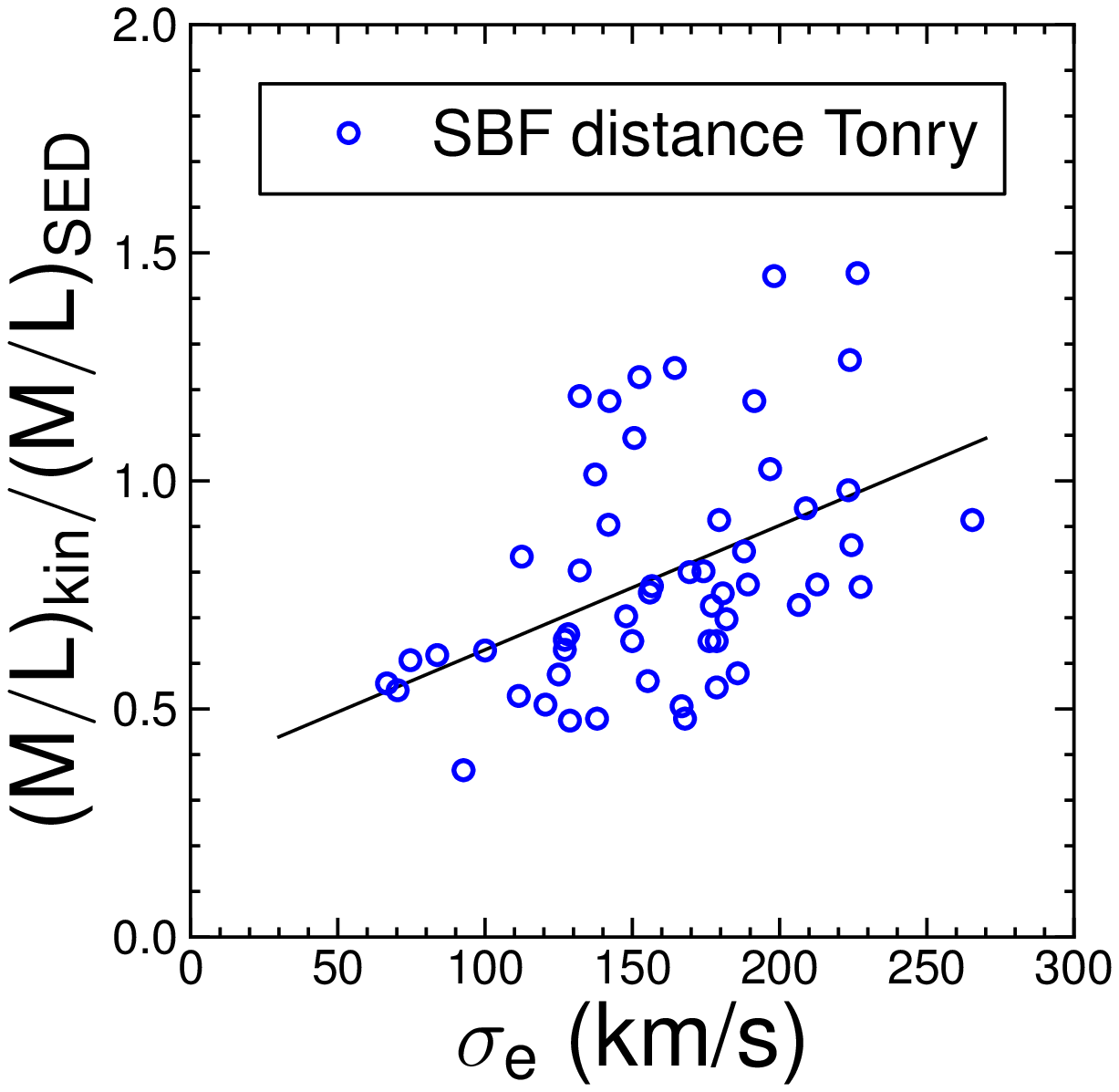}
\caption{The trend of the IMF mismatch parameter $\alpha=(M/L)_{\rm kin}/(M/L)_{\rm SED}$ with effective velocity dispersion for different sub-samples of galaxies. The left and right column show samples selected based on distance and distance measurement method, respectively. The distance selection criterion of 25 Mpc corresponds approximately to the transition from SBF distances to $\rm{V_{hel}}$ distances, avoiding the bias that is introduced by the availability of SBF distance measurements at this distance. The two panels in each row correspond to roughly the same galaxy selections. Top row: most galaxies at $D > 25$ Mpc have a redshift distance determination; middle row: most Virgo galaxies have an SBF distance from \protect\citet{Mei07}; bottom row: most non-Virgo galaxies closer than 25 Mpc have an SBF distance from \protect\citet{Tonry01}. Most of the IMF trend with velocity dispersion comes from the set of non-Virgo galaxies at $D < 25$ Mpc or, equivalently, from the set of galaxies with Tonry SBF distances. This is the same set that shows a distance dependence of the IMF mismatch parameter. Solid lines represent the best fit linear relation for all panels with a Pearson $R^2$ correlation of at least 0.12. The remaining two panels have a Pearson $R^2$ of 0.05. See Table \ref{TableCorrelation} for all corresponding statistics.}
\label{GraphABCDEF}
\end{figure}

The question arises whether this possible bias with distance is in any way related to a possible bias with velocity dispersion, since these appear to be the only two variables that show a systematic trend with the IMF mismatch parameter. Table \ref{TableCorrelation} and Figure \ref{GraphABCDEF} show that this indeed seems to be the case. Exactly the same data set is responsible for most of the correlation of the IMF mismatch parameter $\alpha$ with velocity dispersion as was responsible for most of the correlation between $\alpha$ and distance. For galaxies at distances larger than 25 Mpc there is no clear indication of a systematic IMF variation, nor is there for Virgo galaxies. The systematic trend with velocity dispersion is almost entirely due to the non-Virgo galaxies closer than 25 Mpc. The same trends appear if we select on the three corresponding main distance methods.

There is a striking difference in the IMF trends with velocity dispersion between the two SBF distance sources that are used as input, \citet{Tonry01} and \citet{Mei07}. The SBF method is believed to be the most accurate distance measure for close-by ETGs. The method is based on the assumption that in the observed region the stars sample a homogeneous distribution in space. Fluctuations in brightness are then caused by shot noise. The relative size of these fluctuations contains information about the average number of stars per point spread function. For ETGs that are further away this number of stars will be larger and the relative fluctuations in brightness will be smaller. Although the SBF method can be quite precise, it is an indirect way of measuring distance and may therefore be prone to unknown biases. If all stars would be equally bright then the method would be theoretically simple, but in reality different galaxies consist of different populations of stars, be it because of differences in age, metallicity or possibly the IMF of the galaxy. For this reason the SBF method is calibrated observationally as a function of colour.

The Tonry distance scale is calibrated as a function of (V-I) colour, by comparing with different distance estimates for groups. The Mei distance scale is calibrated as a function of ($g_{475}-z_{850}$) colour. Since the Mei sample consists of galaxies that belong to the Virgo Cluster, the SBF distance is calibrated as a function of colour requiring that different colour galaxies are homogeneously distributed in distance. Table \ref{TableMei} shows the $R^2$ correlation coefficients for the correlations between the spatial distribution of galaxies (in distance, right ascension, declination) and $g_{475}-z_{850}$ colour as well as between spatial distribution and the ATLAS$^{\rm{3D}}$ velocity dispersion. The colour-distance correlation was made to disappear by calibrating the colour-dependent SBF distance, such that the distribution in this direction is as uniform as it is in the transverse directions. Colour and velocity dispersion are highly correlated. Table \ref{TableMei} shows that removing the colour-distance dependence for the Virgo galaxies has also automatically removed the $\sigma_e$-distance dependence.

\begin{table}
\begin{center}
\begin{tabular}{llll}
\hline
 &  Distance & RA & DEC \\   \hline
g(475)-z(850) colour & 0.008 & 0.00007 & 0.03\\  
$\sigma_e$ (ATLAS$^{\rm{3D}}$) & 0.01 & 0.0009 & 0.003\\   \hline
\end{tabular}
\caption{Pearson $R^2$ correlation coefficients between 3D spatial variables of the Virgo galaxies and colour as given by \citet{Mei07} or velocity dispersion as given by \citet{ATLAS15}. Galaxy colours do not correlate significantly with right ascension or declination. The colour - SBF magnitude relation is calibrated by requiring that the same holds in the radial direction. Because colour and $\sigma_e$ are highly correlated, this removes the $\sigma_e$ trend with distance as well.}
\label{TableMei}
\end{center}
\end{table}

\begin{figure*}
\includegraphics[width=1.0\columnwidth]{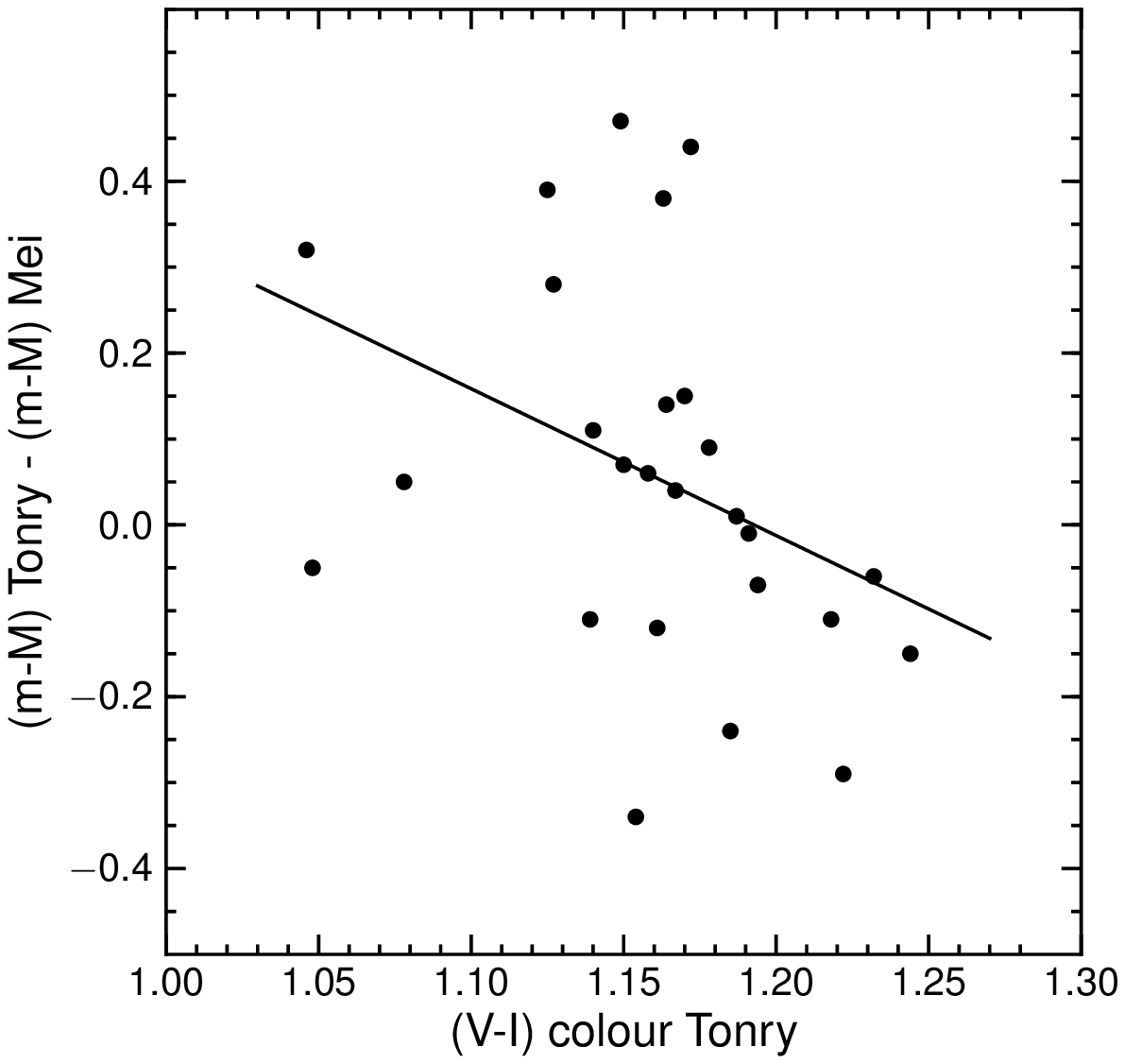}
\includegraphics[width=1.0\columnwidth]{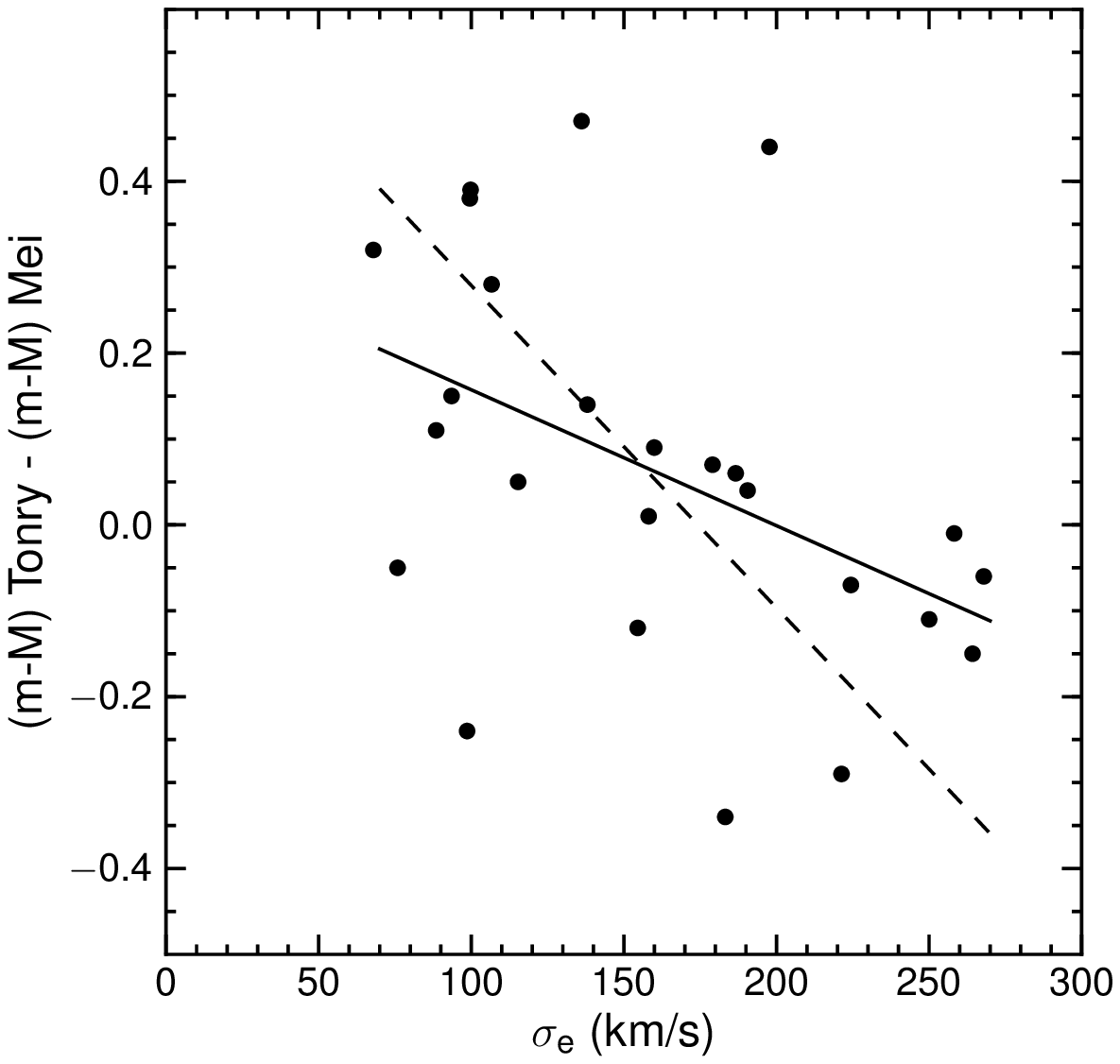}
\caption{For the galaxies that have an SBF distance determination from both \protect\citet{Tonry01} and \protect\cite{Mei07}, the difference in distance modulus is plotted as a function of V-I color from Tonry (left panel) and the effective velocity dispersion from ATLAS$^{\rm{3D}}$ (right panel). A trend in distance calibration is visible in both panels, with respective $R^2$ correlations of 0.14 and 0.20. The dashed line in the right panel represents the systematic bias that would be needed to completely explain the difference in $(\sigma_e,\alpha)$ trend between Tonry and Mei. The observed bias is roughly half of what is needed.}
\label{GraphTonMei}
\end{figure*}

This SBF distance calibration with colour is different for the Tonry dataset. Figure \ref{GraphTonMei} shows the difference in distance modulus for the 26 galaxies that are part of both the \citet{Tonry01} SBF catalog and the \citet{Mei07} SBF catalog. Although one should be cautious in overinterpreting this data due to small number statistics, there are clear trends in the distance difference between the two data sets with both colour and effective velocity dispersion (as determined by ATLAS$^{\rm{3D}}$). For high velocity dispersion the Tonry distance is systematically smaller than the Mei distance and vice versa. This means that for high $\sigma_e$ the JAM method will systematically give a higher IMF mismatch parameter for the Tonry distance than for the Mei distance. This effect is about half of what is needed to fully explain the difference in ($\sigma_e$,$\alpha$) slope in the middle-right and bottom-right panels of Figure \ref{GraphABCDEF}, assuming the same correlation holds for the non-Virgo galaxies that do not have a Mei distance determination.

\begin{figure}
\includegraphics[width=1.0\columnwidth]{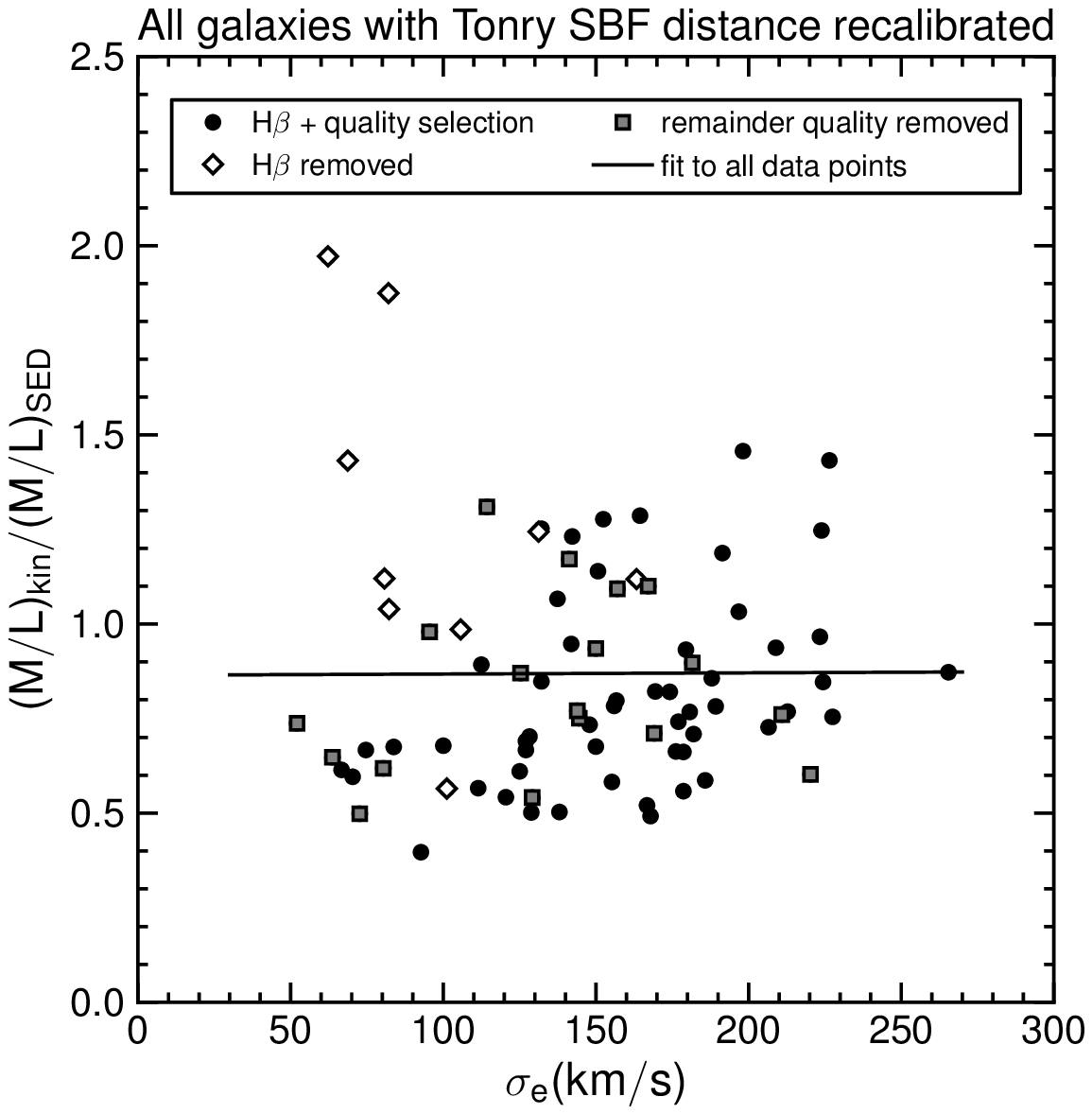}
\caption{Trend of the IMF mismatch parameter $\alpha=(M/L)_{\rm kin}/(M/L)_{\rm SED}$ with effective velocity dispersion for all galaxies that have a \protect\citet{Tonry01} SBF distance in ATLAS$^{\rm{3D}}$. Data points have been recalibrated to account for the $\sigma_e$ dependent difference with \protect\citet{Mei07} SBF distances. The fit to all data points has an $R^2$ of 0.00002 and slope of 0.00003.}
\label{GraphNoCor}
\end{figure}

If we adjust the IMF mismatch parameter with simple scaling relations from Mei to Tonry for the set of galaxies that have a distance determination by both, the best-fit slope of this subset for the ($\sigma_e$,$\alpha$) relation increases from 0.00090 to 0.00140, while the value of $R^2$ increases from 0.221 to 0.338. If we do the opposite for the high quality Tonry galaxies, Figure \ref{GraphABCDEF} (bottom right panel), using $\alpha_{new}=\alpha_{old}\cdot(1.156-7.591 \cdot 10^{-4}\cdot\sigma_e)$,  the best-fit slope decreases from 0.0027 to 0.0022, while the value of $R^2$ decreases from 0.215 to 0.148. The calibration effect is significant, but not sufficient to completely explain the difference in ($\sigma_e$,$\alpha$) trend between the two sets. Qualitatively the conclusion that there is an IMF trend seems to hold. However, one should keep in mind that this conclusion also depends on the $H\beta$ selection effects discussed earlier. For example, the strongest trend of Figure \ref{GraphABCDEF} (bottom right panel) can be made to completely disappear by both recalibrating the Tonry SBF distance and including the ``lower quality JAM fit'' galaxies, see Figure \ref{GraphNoCor}. Moreover, in section \ref{SectionCompleteness} we showed that the effect of mass completeness of the galaxy sample on the inferred ($\sigma_e$,$\alpha$) relation can be large.

\section{Conclusions}
\label{SectionConclusion}

We have asserted the evidence for a non-universal IMF by the ATLAS$^{\rm{3D}}$ Survey and the systematic trend of these IMF variations with the effective velocity dispersion of the Early Type Galaxies.

\begin{itemize}
\item We analysed the correlation between the kinematic mass to light ratio and the IMF mismatch parameter from \citet{Cappellari12}. We show that a similar correlation could arise from Gaussian measurement errors on the kinematic mass to light ratio of the order 30\%, i.e. larger than anticipated (Figures \ref{GraphStarSalp}, \ref{GraphExtra}). However, the observed correlation is somewhat larger than expected from this Gaussian error simulation. The inferred IMF variation hence depends crucially on the precise understanding of the modelling and measurement errors. For this reason, secondary evidence in the form of a large trend of the inferred IMF with another astrophysical variable would be very helpful. The largest trend (at Pearson $R^2=0.11$) is found for velocity dispersion within an effective radius. 
\item Part of the trend of the IMF with velocity dispersion depends on a galaxy selection on $H\beta$ absorption, meant to exclude galaxies with a strong radial gradient in stellar populations (Figures \ref{GraphStarSalp}, \ref{GraphSigmaM}). Although this selection might be unavoidable due to the larger errors in the determinations of $M_{\rm{SED}}$ and $M_{\rm{kin}}$, one should keep in mind the bias that it produces, especially since these are mostly low velocity galaxies with a high IMF mismatch parameter, opposing the trend of the other galaxies. 
\item The IMF trend with velocity dispersion is not accompanied by an IMF trend with mass inferred from SED fitting (Figure \ref{GraphSigmaM}). Thus, contrary to what one might expect \citep{McGee14}, taken at face value, the ATLAS$^{\rm 3D}$ results imply no significant changes in the shape of the observed galaxy stellar mass function (Figure \ref{GraphGSMF}).
\item The ATLAS$^{\rm{3D}}$ Survey is selected to an absolute K-band magnitude $M_{K}$ of $-21.5$. We estimate that this results in incompleteness for masses below $2 \times 10^{10} {\rm {\rm M_{\odot} }}$ (Figure \ref{GraphComp}) (higher than the low-end mass of $\sim 6 \times 10^{9} {\rm M_{\odot}}$ \citep{ATLAS1}). Below this completeness limit the mass plane (MP) as defined by \citet{ATLAS15,ATLAS20} is expected to be affected by completeness effects. The inferred trend between IMF and velocity dispersion is dependent on the precise selection cut-off at the low mass end used in the fit. Specifically, restricting the galaxy sample to the domain $M_{\rm kin}>2 \times 10^{10} {\rm {\rm M_{\odot} }}$ removes the IMF trend with velocity dispersion, whereas the trend is relatively unaffected for a similar sample selection on $M_{\rm SED}$  (Figure \ref{GraphMassConstraint}).
\item Apart from a trend of the IMF mismatch parameter $\alpha=(M/L)_{\rm kin}/(M/L)_{\rm SED}$ with velocity dispersion, we also find a trend with distance (Figure \ref{GraphDistance}). If the correlation between IMF and distance were genuine, then it would presumably be due to environment. However, we find no correlation between the IMF of nearest neighbours. 
\item Selecting galaxies based on the method that was used to measure their distance (distance is used as input in the kinematical fitting procedure) shows that both the IMF trend with distance and the IMF trend with velocity dispersion are concentrated in the subset of galaxies that have a distance determination from \citet{Tonry01} (Figure \ref{GraphABCDEF}). Equivalently, both trends are concentrated in the subset of galaxies that are closer than 25 Mpc and that do not belong to the Virgo Cluster\footnote{The probability of an IMF-velocity dispersion correlation at least as large as that observed for the (non-Virgo, closer than 25 Mpc) galaxy subsample, from a random subsample of galaxies is 1.5\%.}. Most galaxies in the Virgo Cluster have a distance determination from \citet{Mei07}. The subset of galaxies more distant than 25 Mpc shows no IMF trend with velocity dispersion\footnote{The probability of an IMF-velocity dispersion correlation at least as small as that observed for the subsample of galaxies further than 25 Mpc, from a random subsample of galaxies is 12\%.}.
\item Part of the difference in the IMF trend with velocity dispersion between the ETGs with a distance determination from \citet{Tonry01} and those with a distance determination from \citet{Mei07} can be traced back to calibration differences of the SBF distance scale with colour (Figure \ref{GraphTonMei}). The empirical colour-calibration from \citet{Mei07} automatically removes any correlation between distance and velocity dispersion for Virgo galaxies (Table \ref{TableMei}). It also reduces the kinematically deduced IMF trend with velocity dispersion with respect to \citet{Tonry01}. Since this conclusion is reached by comparing the 26 galaxies that have a distance measurement by both \citet{Tonry01} and \citet{Mei07} it might be affected by small number statistics.
\end{itemize}

The dependence of the IMF - $\sigma$ relation on the mass cutoff suggests that it would be valuable to extend the dataset to a lower mass completeness limit (currently at $2\times10^{10} M_{\odot}$). This can rule out the possibility that selection effects contribute to the IMF dependence on velocity dispersion.

This study does not rule out the existence of IMF variations or the correlation of these with velocity dispersion, but it does point out several independent effects that can mimic IMF variations within the framework of the ATLAS$^{\rm{3D}}$ analysis. We need a better understanding and control over random- and systematic errors in ATLAS$^{\rm{3D}}$-like analyses and ultimately we need precision agreement between the different experimental probes of the galactic scale IMF.

\section*{Acknowledgements}

We thank Jarle Brinchmann, Konrad Kuijken, Ivo Labbe and Adam Muzzin for helpful discussions. We thank Sean McGee for a careful reading of the manuscript. We thank the referee for providing constructive comments and help in improving the contents of this paper. We  gratefully acknowledge support from the European Research Council under the European Union's Seventh Framework Programme (FP7/2007-2013) / ERC Grant agreement 278594-GasAroundGalaxies.

\bibliographystyle{mn2e} 
\bibliography{Bibliography}

\begin{thebibliography}{35}
\expandafter\ifx\csname natexlab\endcsname\relax\def\natexlab#1{#1}\fi

\bibitem[{{Bacon} {et~al.}(2001){Bacon}, {Copin}, {Monnet}, {Miller},
  {Allington-Smith}, {Bureau}, {Carollo}, {Davies}, {Emsellem}, {Kuntschner},
  {Peletier}, {Verolme}, \& {de Zeeuw}}]{Bacon01}
{Bacon} R., {Copin} Y., {Monnet} G., {Miller} B.~W., {Allington-Smith} J.~R.,
  {Bureau} M., {Carollo} C.~M., {Davies} R.~L., {Emsellem} E., {Kuntschner} H.,
  {Peletier} R.~F., {Verolme} E.~K., {de Zeeuw} P.~T., 2001, \mnras, 326, 23

\bibitem[{{Barnab{\`e}} {et~al.}(2013){Barnab{\`e}}, {Spiniello}, {Koopmans},
  {Trager}, {Czoske}, \& {Treu}}]{Barnabe13}
{Barnab{\`e}} M., {Spiniello} C., {Koopmans} L.~V.~E., {Trager} S.~C., {Czoske}
  O., {Treu} T., 2013, \mnras, 436, 253

\bibitem[{{Bastian} {et~al.}(2011){Bastian}, {Covey}, \& {Meyer}}]{Bastian11}
{Bastian} N., {Covey} K.~R., {Meyer} M.~R., 2011, in Astronomical Society of
  the Pacific Conference Series, Vol. 448, 16th Cambridge Workshop on Cool
  Stars, Stellar Systems, and the Sun, {Johns-Krull} C., {Browning} M.~K.,
  {West} A.~A., eds., p. 361

\bibitem[{{Brewer} {et~al.}(2012){Brewer}, {Dutton}, {Treu}, {Auger},
  {Marshall}, {Barnab{\`e}}, {Bolton}, {Koo}, \& {Koopmans}}]{Brewer12}
{Brewer} B.~J., {Dutton} A.~A., {Treu} T., {Auger} M.~W., {Marshall} P.~J.,
  {Barnab{\`e}} M., {Bolton} A.~S., {Koo} D.~C., {Koopmans} L.~V.~E., 2012,
  \mnras, 422, 3574

\bibitem[{{Cappellari}(2008)}]{Cappellari08}
{Cappellari} M., 2008, \mnras, 390, 71

\bibitem[{{Cappellari}(2012)}]{Cappellari12*}
---, 2012, ArXiv e-prints

\bibitem[{{Cappellari} {et~al.}(2011{\natexlab{a}}){Cappellari}, {Emsellem},
  {Krajnovi{\'c}}, {McDermid}, {Scott}, {Verdoes Kleijn}, {Young}, {Alatalo},
  {Bacon}, {Blitz}, {Bois}, {Bournaud}, {Bureau}, {Davies}, {Davis}, {de
  Zeeuw}, {Duc}, {Khochfar}, {Kuntschner}, {Lablanche}, {Morganti}, {Naab},
  {Oosterloo}, {Sarzi}, {Serra}, \& {Weijmans}}]{ATLAS1}
{Cappellari} M., {Emsellem} E., {Krajnovi{\'c}} D., {McDermid} R.~M., {Scott}
  N., {Verdoes Kleijn} G.~A., {Young} L.~M., {Alatalo} K., {Bacon} R., {Blitz}
  L., {Bois} M., {Bournaud} F., {Bureau} M., {Davies} R.~L., {Davis} T.~A., {de
  Zeeuw} P.~T., {Duc} P.-A., {Khochfar} S., {Kuntschner} H., {Lablanche} P.-Y.,
  {Morganti} R., {Naab} T., {Oosterloo} T., {Sarzi} M., {Serra} P., {Weijmans}
  A.-M., 2011{\natexlab{a}}, \mnras, 413, 813

\bibitem[{{Cappellari} {et~al.}(2011{\natexlab{b}}){Cappellari}, {Emsellem},
  {Krajnovi{\'c}}, {McDermid}, {Serra}, {Alatalo}, {Blitz}, {Bois}, {Bournaud},
  {Bureau}, {Davies}, {Davis}, {de Zeeuw}, {Khochfar}, {Kuntschner},
  {Lablanche}, {Morganti}, {Naab}, {Oosterloo}, {Sarzi}, {Scott}, {Weijmans},
  \& {Young}}]{ATLAS7}
{Cappellari} M., {Emsellem} E., {Krajnovi{\'c}} D., {McDermid} R.~M., {Serra}
  P., {Alatalo} K., {Blitz} L., {Bois} M., {Bournaud} F., {Bureau} M., {Davies}
  R.~L., {Davis} T.~A., {de Zeeuw} P.~T., {Khochfar} S., {Kuntschner} H.,
  {Lablanche} P.-Y., {Morganti} R., {Naab} T., {Oosterloo} T., {Sarzi} M.,
  {Scott} N., {Weijmans} A.-M., {Young} L.~M., 2011{\natexlab{b}}, \mnras, 416,
  1680

\bibitem[{{Cappellari} {et~al.}(2012){Cappellari}, {McDermid}, {Alatalo},
  {Blitz}, {Bois}, {Bournaud}, {Bureau}, {Crocker}, {Davies}, {Davis}, {de
  Zeeuw}, {Duc}, {Emsellem}, {Khochfar}, {Krajnovi{\'c}}, {Kuntschner},
  {Lablanche}, {Morganti}, {Naab}, {Oosterloo}, {Sarzi}, {Scott}, {Serra},
  {Weijmans}, \& {Young}}]{Cappellari12}
{Cappellari} M., {McDermid} R.~M., {Alatalo} K., {Blitz} L., {Bois} M.,
  {Bournaud} F., {Bureau} M., {Crocker} A.~F., {Davies} R.~L., {Davis} T.~A.,
  {de Zeeuw} P.~T., {Duc} P.-A., {Emsellem} E., {Khochfar} S., {Krajnovi{\'c}}
  D., {Kuntschner} H., {Lablanche} P.-Y., {Morganti} R., {Naab} T., {Oosterloo}
  T., {Sarzi} M., {Scott} N., {Serra} P., {Weijmans} A.-M., {Young} L.~M.,
  2012, \nat, 484, 485

\bibitem[{{Cappellari} {et~al.}(2013{\natexlab{a}}){Cappellari}, {McDermid},
  {Alatalo}, {Blitz}, {Bois}, {Bournaud}, {Bureau}, {Crocker}, {Davies},
  {Davis}, {de Zeeuw}, {Duc}, {Emsellem}, {Khochfar}, {Krajnovi{\'c}},
  {Kuntschner}, {Morganti}, {Naab}, {Oosterloo}, {Sarzi}, {Scott}, {Serra},
  {Weijmans}, \& {Young}}]{ATLAS20}
{Cappellari} M., {McDermid} R.~M., {Alatalo} K., {Blitz} L., {Bois} M.,
  {Bournaud} F., {Bureau} M., {Crocker} A.~F., {Davies} R.~L., {Davis} T.~A.,
  {de Zeeuw} P.~T., {Duc} P.-A., {Emsellem} E., {Khochfar} S., {Krajnovi{\'c}}
  D., {Kuntschner} H., {Morganti} R., {Naab} T., {Oosterloo} T., {Sarzi} M.,
  {Scott} N., {Serra} P., {Weijmans} A.-M., {Young} L.~M., 2013{\natexlab{a}},
  \mnras, 432, 1862

\bibitem[{{Cappellari} {et~al.}(2013{\natexlab{b}}){Cappellari}, {Scott},
  {Alatalo}, {Blitz}, {Bois}, {Bournaud}, {Bureau}, {Crocker}, {Davies},
  {Davis}, {de Zeeuw}, {Duc}, {Emsellem}, {Khochfar}, {Krajnovi{\'c}},
  {Kuntschner}, {McDermid}, {Morganti}, {Naab}, {Oosterloo}, {Sarzi}, {Serra},
  {Weijmans}, \& {Young}}]{ATLAS15}
{Cappellari} M., {Scott} N., {Alatalo} K., {Blitz} L., {Bois} M., {Bournaud}
  F., {Bureau} M., {Crocker} A.~F., {Davies} R.~L., {Davis} T.~A., {de Zeeuw}
  P.~T., {Duc} P.-A., {Emsellem} E., {Khochfar} S., {Krajnovi{\'c}} D.,
  {Kuntschner} H., {McDermid} R.~M., {Morganti} R., {Naab} T., {Oosterloo} T.,
  {Sarzi} M., {Serra} P., {Weijmans} A.-M., {Young} L.~M., 2013{\natexlab{b}},
  \mnras, 432, 1709

\bibitem[{{Chabrier}(2003)}]{Chabrier03}
{Chabrier} G., 2003, \pasp, 115, 763

\bibitem[{{Conroy} {et~al.}(2013){Conroy}, {Dutton}, {Graves}, {Mendel}, \&
  {van Dokkum}}]{Conroy13}
{Conroy} C., {Dutton} A.~A., {Graves} G.~J., {Mendel} J.~T., {van Dokkum}
  P.~G., 2013, \apjl, 776, L26

\bibitem[{{Conroy} \& {van Dokkum}(2012)}]{Conroy12}
{Conroy} C., {van Dokkum} P.~G., 2012, \apj, 760, 71

\bibitem[{{Dutton} {et~al.}(2013){Dutton}, {Macci{\`o}}, {Mendel}, \&
  {Simard}}]{Dutton13}
{Dutton} A.~A., {Macci{\`o}} A.~V., {Mendel} J.~T., {Simard} L., 2013, \mnras,
  432, 2496

\bibitem[{{Emsellem} {et~al.}(1994){Emsellem}, {Monnet}, \&
  {Bacon}}]{Emsellem94}
{Emsellem} E., {Monnet} G., {Bacon} R., 1994, \aap, 285, 723

\bibitem[{{Kirk} \& {Myers}(2011)}]{Kirk11}
{Kirk} H., {Myers} P.~C., 2011, \apj, 727, 64

\bibitem[{{Kroupa} {et~al.}(2013){Kroupa}, {Weidner}, {Pflamm-Altenburg},
  {Thies}, {Dabringhausen}, {Marks}, \& {Maschberger}}]{Kroupa13}
{Kroupa} P., {Weidner} C., {Pflamm-Altenburg} J., {Thies} I., {Dabringhausen}
  J., {Marks} M., {Maschberger} T., 2013, {The Stellar and Sub-Stellar Initial
  Mass Function of Simple and Composite Populations}, {Oswalt} T.~D., {Gilmore}
  G., eds., p. 115

\bibitem[{{La Barbera} {et~al.}(2013){La Barbera}, {Ferreras}, {Vazdekis}, {de
  la Rosa}, {de Carvalho}, {Trevisan}, {Falc{\'o}n-Barroso}, \&
  {Ricciardelli}}]{Barbera13}
{La Barbera} F., {Ferreras} I., {Vazdekis} A., {de la Rosa} I.~G., {de
  Carvalho} R.~R., {Trevisan} M., {Falc{\'o}n-Barroso} J., {Ricciardelli} E.,
  2013, \mnras, 433, 3017

\bibitem[{{Mart{\'{\i}}n-Navarro} {et~al.}(2014){Mart{\'{\i}}n-Navarro}, {La
  Barbera}, {Vazdekis}, {Falc{\'o}n-Barroso}, \& {Ferreras}}]{MartinNavarro14}
{Mart{\'{\i}}n-Navarro} I., {La Barbera} F., {Vazdekis} A.,
  {Falc{\'o}n-Barroso} J., {Ferreras} I., 2014, ArXiv e-prints

\bibitem[{{McDermid} {et~al.}(2014){McDermid}, {Cappellari}, {Alatalo},
  {Bayet}, {Blitz}, {Bois}, {Bournaud}, {Bureau}, {Crocker}, {Davies}, {Davis},
  {de Zeeuw}, {Duc}, {Emsellem}, {Khochfar}, {Krajnovi{\'c}}, {Kuntschner},
  {Morganti}, {Naab}, {Oosterloo}, {Sarzi}, {Scott}, {Serra}, {Weijmans}, \&
  {Young}}]{McDermid14}
{McDermid} R.~M., {Cappellari} M., {Alatalo} K., {Bayet} E., {Blitz} L., {Bois}
  M., {Bournaud} F., {Bureau} M., {Crocker} A.~F., {Davies} R.~L., {Davis}
  T.~A., {de Zeeuw} P.~T., {Duc} P.-A., {Emsellem} E., {Khochfar} S.,
  {Krajnovi{\'c}} D., {Kuntschner} H., {Morganti} R., {Naab} T., {Oosterloo}
  T., {Sarzi} M., {Scott} N., {Serra} P., {Weijmans} A.-M., {Young} L.~M.,
  2014, \apjl, 792, L37

\bibitem[{{McGee} {et~al.}(2014){McGee}, {Goto}, \& {Balogh}}]{McGee14}
{McGee} S.~L., {Goto} R., {Balogh} M.~L., 2014, \mnras, 438, 3188

\bibitem[{{Mei} {et~al.}(2007){Mei}, {Blakeslee}, {C{\^o}t{\'e}}, {Tonry},
  {West}, {Ferrarese}, {Jord{\'a}n}, {Peng}, {Anthony}, \& {Merritt}}]{Mei07}
{Mei} S., {Blakeslee} J.~P., {C{\^o}t{\'e}} P., {Tonry} J.~L., {West} M.~J.,
  {Ferrarese} L., {Jord{\'a}n} A., {Peng} E.~W., {Anthony} A., {Merritt} D.,
  2007, \apj, 655, 144

\bibitem[{{Mould} {et~al.}(2000){Mould}, {Huchra}, {Freedman}, {Kennicutt},
  {Ferrarese}, {Ford}, {Gibson}, {Graham}, {Hughes}, {Illingworth}, {Kelson},
  {Macri}, {Madore}, {Sakai}, {Sebo}, {Silbermann}, \& {Stetson}}]{Mould00}
{Mould} J.~R., {Huchra} J.~P., {Freedman} W.~L., {Kennicutt} Jr. R.~C.,
  {Ferrarese} L., {Ford} H.~C., {Gibson} B.~K., {Graham} J.~A., {Hughes}
  S.~M.~G., {Illingworth} G.~D., {Kelson} D.~D., {Macri} L.~M., {Madore} B.~F.,
  {Sakai} S., {Sebo} K.~M., {Silbermann} N.~A., {Stetson} P.~B., 2000, \apj,
  529, 786

\bibitem[{{Moustakas} {et~al.}(2013){Moustakas}, {Coil}, {Aird}, {Blanton},
  {Cool}, {Eisenstein}, {Mendez}, {Wong}, {Zhu}, \& {Arnouts}}]{Moustakas13}
{Moustakas} J., {Coil} A.~L., {Aird} J., {Blanton} M.~R., {Cool} R.~J.,
  {Eisenstein} D.~J., {Mendez} A.~J., {Wong} K.~C., {Zhu} G., {Arnouts} S.,
  2013, \apj, 767, 50

\bibitem[{{Oguri} {et~al.}(2014){Oguri}, {Rusu}, \& {Falco}}]{Oguri14}
{Oguri} M., {Rusu} C.~E., {Falco} E.~E., 2014, \mnras

\bibitem[{{Pastorello} {et~al.}(2014){Pastorello}, {Forbes}, {Foster},
  {Brodie}, {Usher}, {Romanowsky}, {Strader}, \& {Arnold}}]{Pastorello14}
{Pastorello} N., {Forbes} D.~A., {Foster} C., {Brodie} J.~P., {Usher} C.,
  {Romanowsky} A.~J., {Strader} J., {Arnold} J.~A., 2014, \mnras, 442, 1003

\bibitem[{{Peacock} {et~al.}(2014){Peacock}, {Zepf}, {Maccarone}, {Kundu},
  {Gonzalez}, {Lehmer}, \& {Maraston}}]{Peacock14}
{Peacock} M.~B., {Zepf} S.~E., {Maccarone} T.~J., {Kundu} A., {Gonzalez} A.~H.,
  {Lehmer} B.~D., {Maraston} C., 2014, ArXiv e-prints

\bibitem[{{Smith}(2014)}]{Smith14}
{Smith} R.~J., 2014, ArXiv e-prints

\bibitem[{{Smith} \& {Lucey}(2013)}]{Smith13}
{Smith} R.~J., {Lucey} J.~R., 2013, \mnras, 434, 1964

\bibitem[{{Tonry} {et~al.}(2001){Tonry}, {Dressler}, {Blakeslee}, {Ajhar},
  {Fletcher}, {Luppino}, {Metzger}, \& {Moore}}]{Tonry01}
{Tonry} J.~L., {Dressler} A., {Blakeslee} J.~P., {Ajhar} E.~A., {Fletcher}
  A.~B., {Luppino} G.~A., {Metzger} M.~R., {Moore} C.~B., 2001, \apj, 546, 681

\bibitem[{{Tortora} {et~al.}(2013){Tortora}, {Romanowsky}, \&
  {Napolitano}}]{Tortora13}
{Tortora} C., {Romanowsky} A.~J., {Napolitano} N.~R., 2013, \apj, 765, 8

\bibitem[{{Treu} {et~al.}(2010){Treu}, {Auger}, {Koopmans}, {Gavazzi},
  {Marshall}, \& {Bolton}}]{Treu10}
{Treu} T., {Auger} M.~W., {Koopmans} L.~V.~E., {Gavazzi} R., {Marshall} P.~J.,
  {Bolton} A.~S., 2010, \apj, 709, 1195

\bibitem[{{van Dokkum} \& {Conroy}(2012)}]{Dokkum12}
{van Dokkum} P.~G., {Conroy} C., 2012, \apj, 760, 70

\bibitem[{{Vazdekis} {et~al.}(2012){Vazdekis}, {Ricciardelli}, {Cenarro},
  {Rivero-Gonz{\'a}lez}, {D{\'{\i}}az-Garc{\'{\i}}a}, \&
  {Falc{\'o}n-Barroso}}]{Vazdekis12}
{Vazdekis} A., {Ricciardelli} E., {Cenarro} A.~J., {Rivero-Gonz{\'a}lez} J.~G.,
  {D{\'{\i}}az-Garc{\'{\i}}a} L.~A., {Falc{\'o}n-Barroso} J., 2012, \mnras,
  424, 157

\end{thebibliography}

\label{lastpage}
\end{document}